\documentclass[journal]{IEEEtran}
\usepackage{amsmath}
\usepackage{amssymb}
\usepackage{amsthm}
\usepackage{amsfonts}
\usepackage{epsfig}
\usepackage{psfrag}
\usepackage{graphicx}
\usepackage{mathrsfs}
\usepackage{color}
\usepackage{cite}
\usepackage[colorlinks=true,linkcolor=black,anchorcolor=black,citecolor=black,filecolor=black,menucolor=black,runcolor=black,urlcolor=black]{hyperref}
\usepackage{multirow}

\usepackage{physics} 
\usepackage{empheq}
\usepackage{authblk}
\usepackage[normalem]{ulem} 
\usepackage{mathtools} 

\usepackage{algorithm}
\usepackage{algpseudocode}

\ifCLASSOPTIONcompsoc
\usepackage[caption=false,font=normalsize,labelfon
t=sf,textfont=sf]{subfig}
\else
\usepackage[caption=false,font=footnotesize]{subfig}
\fi

\usepackage{stfloats}

\theoremstyle{definition}
\newtheorem{assumption}{Assumption}
\newtheorem{proposition}{Proposition}

\newtheorem{lemma}{Lemma}
\newtheorem{remark}{Remark}
\newtheorem{corollary}{Corollary}

\title{Toward Ambient Intelligence: Federated Edge Learning with Task-Oriented Sensing, Computation, and Communication Integration}
\author{Peixi Liu, Guangxu Zhu, \textit{Member, IEEE},  Shuai Wang, \textit{Member, IEEE}, Wei Jiang, \textit{Member, IEEE}, Wu Luo, \textit{Member, IEEE}, H. Vincent Poor, \textit{Life Fellow, IEEE}, Shuguang Cui, \textit{Fellow, IEEE},\vspace{-0.5cm}   
\thanks{Peixi Liu is with State Key Laboratory of Advanced Optical Communication Systems and Networks, School of Electronics, Peking University, Beijing, China, and Shenzhen Research Institute of Big Data, Shenzhen, China (e-mail: liupeixi@pku.edu.cn). (Corresponding authors: Guangxu Zhu, Wei Jiang)} 
\thanks{Guangxu Zhu is with Shenzhen Research Institute of Big Data, Shenzhen, China (e-mail: gxzhu@sribd.cn).}
\thanks{Shuai Wang is with the Guangdong-Hong Kong-Macao Joint Laboratory of Human-Machine Intelligence-Synergy Systems, Shenzhen Institute of Advanced Technology, Chinese Academy of Sciences, Shenzhen 518055, China, and also with the Guangdong Laboratory of Artificial Intelligence and Digital Economy (SZ), Shenzhen 518060, China (e-mail: s.wang@siat.ac.cn).}
\thanks{Wei Jiang and Wu Luo are with State Key Laboratory of Advanced Optical Communication Systems and Networks, School of Electronics, Peking University, Beijing, China (e-mail: jiangwei, luow@pku.edu.cn).} 
\thanks{H. Vincent Poor is with Department of Electrical and Computer Engineering, Princeton University, Princeton, NJ08544, USA (email: poor@princeton.edu).}
\thanks{Shuguang Cui is with the School of Science and Engineering (SSE), the Future Network of Intelligence Institute (FNii), and the Guangdong Provincial Key Laboratory of Future Networks of Intelligence, The Chinese University of Hong Kong (Shenzhen), and Shenzhen Research Institute of Big Data, Shenzhen, China. He is also with Peng Cheng Laboratory (e-mail: shuguangcui@cuhk.edu.cn).}
}
\begin{document}

%
%
	\maketitle
\begin{abstract}
	With the breakthroughs in deep learning and contactless sensors, the recent years have witnessed a rise of ambient intelligence applications and services, spanning from healthcare delivery to intelligent home.
	\textit{Federated edge learning} (FEEL), as a privacy-enhancing paradigm of collaborative learning at the network edge, is expected to be the core engine to achieve ambient intelligence.
	\textit{Sensing, computation, and communication} (SC$^{2}$) are highly coupled processes in FEEL and need to be jointly designed in a task-oriented manner for pursuing the best FEEL performance under the stringent resource constraints at the edge devices.  
	However, this remains an open problem as there is a lack of theoretical understanding on how the SC$^{2}$ resource jointly affect the FEEL performance. 
	In this paper, we address the problem of joint SC$^{2}$ resource allocation for FEEL via a concrete case study of human motion recognition based on wireless sensing in ambient intelligence. 
	First, by analyzing the wireless sensing process in human motion recognition, we find that there exists a thresholding value for the sensing transmit power, exceeding which yields sensing data samples with approximately the same satisfactory quality. 
	Then, the joint SC$^{2}$ resource allocation problem is cast to maximize the convergence speed of FEEL, under the constraints on training time, energy supply, and sensing quality of each edge device. 
	Solving this problem entails solving two subproblems in order: the first one reduces to determine the joint sensing and communication resource allocation that maximizes the total number of samples that can be sensed during the entire training process; the second one concerns the partition of the attained total number of sensed samples over all the communication rounds to determine the batch size at each round for convergence speed maximization.
	The first subproblem on joint sensing and communication resource allocation is converted to a single-variable optimization problem by exploiting the derived relation between different control variables (resources), which thus allows an efficient solution via one-dimensional grid search. 
	For the second subproblem, it is found that the number of samples to be sensed (or batch size) at each round is a decreasing function of the loss function value attained at the round.
	Based on this relationship, the approximate optimal batch size at each communication round is derived in closed-form as a function of the round index. 
	Finally, extensive simulation results are provided to validate the superiority of the proposed joint SC$^{2}$ resource allocation scheme over baseline schemes in terms of FEEL performance.
\end{abstract}


\section{Introduction}
With the continuous integration of information and communication technologies and the all-round penetration of \textit{artificial intelligence} (AI), we can envision that the future 6G network will no longer serve only the single purpose of data transmission, but will need to support connected intelligence services and ubiquitous intelligence applications\cite{Letaief2021Edge-6G,Saad2019Netw_6G}.
One emerging technology of ubiquitous intelligence, called \textit{ambient intelligence}, has shown significant potential in improving the efficiency of healthcare delivery and quality of life \cite{Haque2020Nature}.
The goal of ambient intelligence is to build physical spaces that are sensitive and responsive to the inputs triggered by humans and to provide low-latency, high-accurate, scalable, and resilient services with the help of AI technologies and contactless sensors \cite{Dunne2021ACM-ambient}.
Therefore, in order to support ambient intelligence, the future 6G network needs to undergo a paradigm shift from a pure information delivery pipeline to an integrated sensing-computation-communication multi-functional platform.

Demand for low-latency in ambient intelligence has prompted the development of AI technologies at the network edge, leading to an emerging research area known as \textit{edge learning} \cite{Zhu2020CM_FEEL}, which advocates local data processing at the edge and thus avoids the long delay resulting from the data transfer from the edge to the remote cloud server.
Specifically, edge learning allows the edge devices to process the sensory data from their onboard sensors and makes full use of the computing power at the edge to learn AI models customized for specific tasks.
Among the techniques in edge learning, \textit{federated edge learning} (FEEL) is a popular collaborative distributed learning paradigm that trains a global \textit{machine learning} (ML) model over wireless networks while helping to preserve data privacy \cite{Zhu2019TWC_broadband,Chen2021JSAC-FEEL,Wang2022arXiv}. 
In a typical training iteration of FEEL, a dedicated edge server first broadcasts the global ML model to the participating edge devices; next, the edge devices compute their respective model-updates using their local data and then upload them to the edge server for further aggregation and global model updating.

\begin{figure}[!t]
	\centering
	\includegraphics[width=0.36\textwidth]{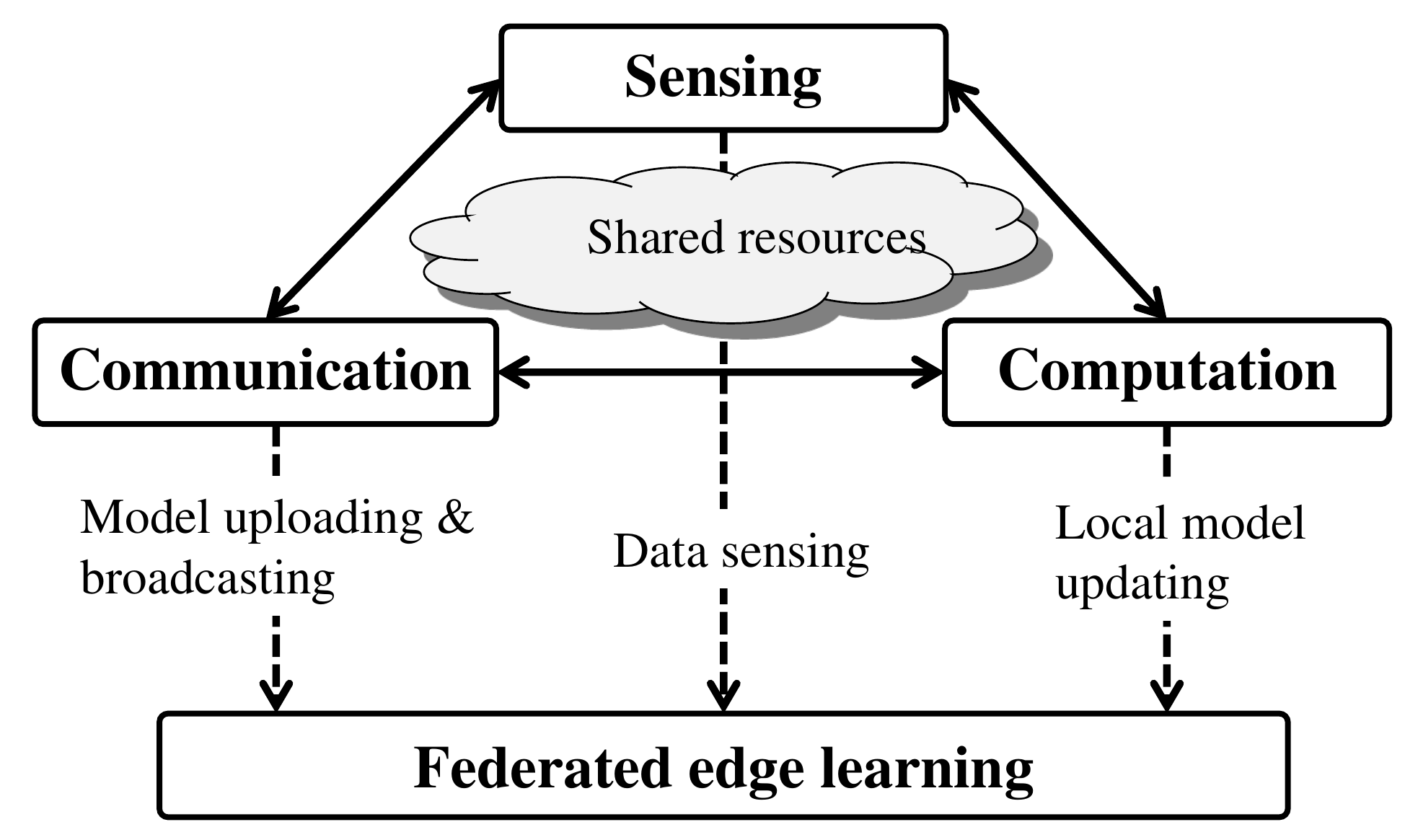}\vspace{-0.2cm}
	\caption{The coupling relationship between sensing, computation, and communication (SC$^{2}$) in FEEL.}\vspace{-0.5cm}
	\label{fig:fig-ISCC}
\end{figure}

Since the ML model is transmitted over wireless links which could introduce training errors due to the limited radio resources (e.g., bandwidth), prior works on FEEL mainly focus on accelerating the learning process from the resource allocation perspective.
For example, the work in \cite{Yang2020TWC-aircomp} optimized the receive beamforming at the edge server to maximize the number of participated edge devices at each iteration of FEEL under peak power constraints.
The works in \cite{Cao2021JSAC-aircomp,Wei2022TCCN-noisy,Amiri2020TWC-fading} designed the power allocation of each edge device to maximize the convergence speed of FEEL under a total energy constraint. 
Besides power and energy constraints, the authors in \cite{Chen2020TWC-joint} and \cite{Guo2021JSTSP-dynamic} also took the latency constraint into consideration, and studied the problem of power allocation and device scheduling in FEEL.
While interesting, the works in \cite{Zhu2019TWC_broadband,Chen2021JSAC-FEEL,Yang2020TWC-aircomp,Cao2021JSAC-aircomp,Wei2022TCCN-noisy,Amiri2020TWC-fading,Chen2020TWC-joint,Guo2021JSTSP-dynamic} all took the learning error as the objective function. Instead, the work in \cite{Yang2020TWC-Energy} minimized total energy consumption under a latency constraint and learning performance guarantee.
Last but not least, from a more practical consideration, the authors in \cite{Wadu2021TC-imCSI} studied resource block allocation under imperfect channel state information.
These prior works mainly focused on the communication and/or computation perspectives and assumed that the data used for training are readily available without considering the data sensing process, which could significantly affect the learning performance as sensing also competes with communication and computation for resources.
As shown in Fig. \ref{fig:fig-ISCC}, \textit{sensing, computation, and communication} (SC$^{2}$) are highly coupled in FEEL and thus need to be seamlessly integrated in a joint design to fully unleash the potential of FEEL. This thus calls for task-oriented SC$^2$ resource management that takes the FEEL performance as the ultimate objective in the resulting optimization.

Among various contactless sensors used in ambient intelligence, wireless sensors are widely adopted inside office buildings, homes, and hospitals due to their unique advantages such as safety, reliability, portability, and affordability \cite{Gurbuz2019SPM_SensingDL}.
Wireless sensors sense the environmental information by analyzing the raw reflected radio signals from the sensing targets, and achieve the functions of detection, ranging, positioning, imaging, and so on.
The fact that radio signals can not only be used for wireless communication but also for environmental sensing is particularly appealing as it allows the same radio hardware to be reused for dual functions. This has led to an increasingly popular research area known as \textit{integrated sensing and communication} (ISAC) \cite{Liu2022JSAC-ISAC-overview,Yuan2021JSTSP-ISAC}. The integration of both functions enjoys the benefit of a smaller form factor for the ISAC devices due to the shared hardware and better management of the shared radio resources, such as power and bandwidth. In view of its promising potential, ISAC is expected to support various ambient intelligence applications, ranging from indoor health monitoring to extended human sensing \cite{Cui2021Netw_ISAC}.
In these applications, fundamental differences in the data domain have driven the development of unique AI-based signal processing. 
In particular, sensing data is not inherently acquired as images, but mostly as complex time series with amplitudes and phases related to the electromagnetic scattering and kinematics of the sensed targets \cite{Nirmal2021CST-radio-sensing}.
Hence, some preprocessing stages, e.g., filtering and time-frequency analysis, are required to transform the raw signals into the data samples as the inputs of ML models.
Since the raw reflected signals are typically mixed with  scatterings from the environment, it poses a significant challenge to generate data samples of approximately the same satisfactory quality over time, which is desirable for training a ML model.

Despite the fact that wireless sensor outputs may not include photos or video, some wireless sensing applications can nevertheless create enormous volumes of data that may contain private information about the users, such as the number of occupants, health conditions, and room sizes. The wealth of sensing data at the network edge should also be leveraged without under fear of privacy leakage. Fortunately, FEEL as a privacy-enhancing learning framework can be applied with wireless sensing in ambient intelligence \cite{Hernandez2021IoT-WiFed}.
In this work, we investigate the joint SC$^{2}$ resource allocation in FEEL by considering a concrete case study, i.e., human motion recognition. 
Specifically, we consider an ISAC-based FEEL system in which multiple ISAC devices obtain their own local datasets for human motion recognition by wireless sensing, and communicate with the server to exchange model-updates over wireless channels. Then  each ISAC device updates its local model using only the latest sensed data samples, the number of which gives its batch size for the current round.
Under this online FEEL setting, the SC$^2$ resources include the transmit power and time for sensing and communication, as well as the batch size to be computed at each round.
In this setting, we address two challenges for the considered system:
1) how to generate data samples with approximately the same satisfactory quality for FEEL over time by wireless sensing;
2) how to jointly allocate the SC$^2$ resources in an task-oriented manner so as to yield the best learning performance.
The findings and the contributions of this work are as follows:
\begin{itemize}
	\item \textbf{Sensing quality analysis:} We present the pipeline of wireless sensing in human motion recognition, empirically analyze the sensing process, and find that the sensing quality does not improve much when the sensing transmit power exceeds a threshold value. This finding suggests that it is sufficient to sense with the said threshold power value for generating data samples of approximately the same satisfactory quality over time, which addresses the first challenge above.
	\item \textbf{Problem formulation and splitting for joint SC$^{2}$ resource allocation:} We formulate the joint SC$^{2}$ resource allocation problem in FEEL, which aims to maximize the convergence speed under total latency and energy constraints. To solve this challenging non-convex problem, we split it into two subproblems without loss of optimality: the one concerns sensing and communication resource allocation to maximize the number of total sensed data samples in the training process; the other concerns the partition of the total sensed data samples over rounds to determine the batch size to be computed at each round for convergence speed maximization. 
	\item \textbf{Optimal joint SC$^{2}$ resource allocation:} To tackle the first subproblem of joint sensing and communication resource allocation, we first analyze the coupled relations between different control variables, by exploiting which, the challenging multi-variable optimization problem is transformed to a single variable problem. This thus allows an efficient solution via one-dimensional grid search. 
	The second subproblem cannot be solved directly due to the lack of an explicit objective function. Instead, we derive an approximate solution in closed-form for the optimal batch sizes at different rounds, based on a theoretical finding that the optimal batch size should be adaptive and increase as the training loss value decreases in the course of training. The solutions of the two subproblems together address the second challenge above.
	\item \textbf{Performance evaluation:} We conduct extensive simulations based on a concrete wireless sensing task of human motion recognition over a high-fidelity wireless sensing simulator \cite{Li2021SPAWC-sensing-model} to evaluate the performance of our proposed joint SC$^{2}$ resource allocation scheme. The superiority of the proposed scheme over other baseline joint SC$^{2}$ resource management schemes is demonstrated. 
\end{itemize} 
%
%
%
%
%
%

\begin{figure}[!t]
	\centering
	\includegraphics[width=0.4\textwidth]{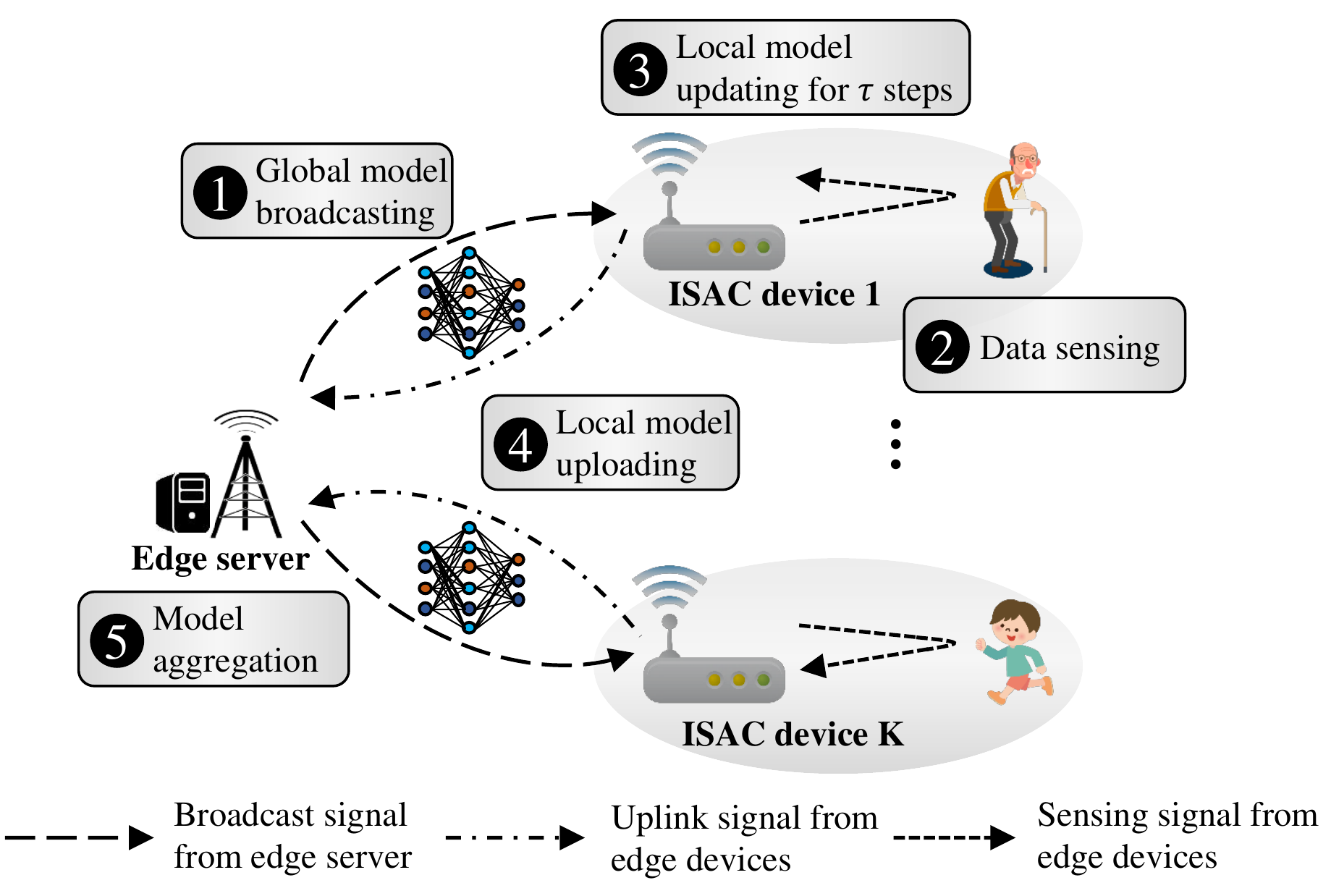}
	\caption{Federated edge learning system with integrated sensing and communication.}\vspace{-0.5cm}
	\label{fig:fig-system-model}
\end{figure}

The remainder of this paper is organized as follows. Sec. \ref{sec:system-model} presents the system model, including FEEL specification, communication and sensing models. Sec. \ref{sec:analysis} analyzes the communication rate and sensing quality in this setting. Sec. \ref{sec:problem-formulation} formulates the SC$^2$ resource allocation problem  of FEEL under total latency and energy constraints. The optimal power allocation of communication and sensing and batch size updating policy are characterized in Sec. \ref{sec:solution}. Simulation results are given in Sec. \ref{sec:simulation}, followed by conclusions in Sec. \ref{sec:conclusion}.


\section{System Model}\label{sec:system-model}

\subsection{Federated Edge Learning}\label{sec:FedAvg}

\begin{figure}[!t]
	\centering
	\includegraphics[width=0.39\textwidth]{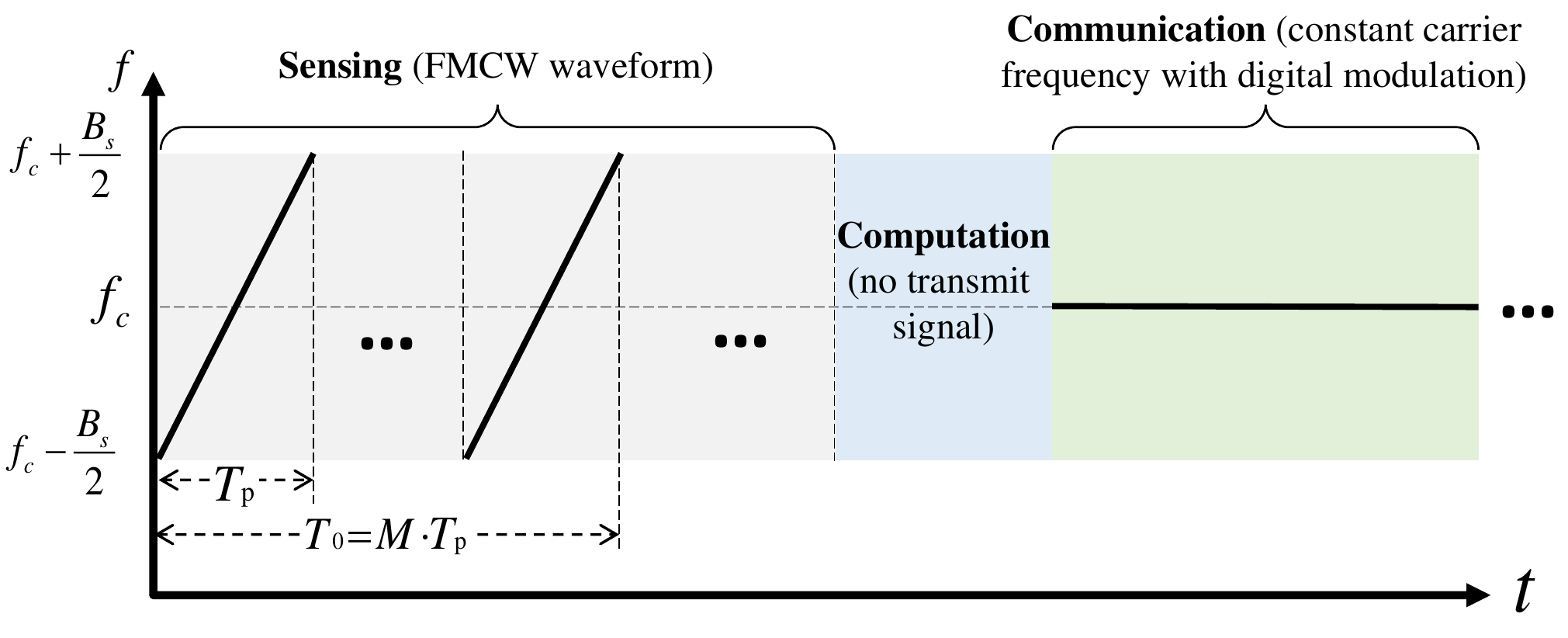}
	\vspace{-0.4cm}
	\caption{Time-frequency diagram of the considered time-switching ISAC waveforms in a communication round.}
	\label{fig:fig-sensing-signal-format}
	\vspace{-0.4cm}
\end{figure}

As shown in Fig. \ref{fig:fig-system-model}, we consider an ISAC-assisted FEEL system comprising $K$ ISAC devices, denoted by a set $\mathcal{K}=\lbrace 1,2,...,K \rbrace$, and a single edge server. \color{black}
Each ISAC device is equipped with a single-antenna ISAC transceiver that can switch between the sensing mode and communication mode as needed in a time-devision manner using a shared radio-frequency front-end circuit\footnote{A practical implementation of such an ISAC device via software-defined radio platform has been demonstrated in \cite{Han2013ISAC-hardware}.} \cite{Han2013ISAC-hardware}. Specifically, in the sensing mode, dedicated radar waveform known as \textit{frequency-modulated continuous-wave} (FMCW) consisting of multiple up-chirps is transmitted \cite{Gurbuz2019SPM_SensingDL}. Then, by processing the received radar echo signals, sensing data that contain the motion information of the human target can be attained at ISAC devices (see Sec. \ref{sec:sensing-quality} for more details of the sensing processing). On the other hand, in the communication mode, constant-frequency carrier modulated by communication data using digital modulation scheme (e.g., QAM) is transmitted. This mode is used for the necessary information exchange between the ISAC device and edge server in the course of FEEL as elaborated in Sec. \ref{sec:comm-model}. For better illustration, the time-frequency diagram for the considered time-switching ISAC waveforms is depicted in Fig. \ref{fig:fig-sensing-signal-format}. \color{black} 
Besides, each ISAC device is equipped with certain computation power and can perform local computations for training and inference of ML models, e.g., \textit{convolutional neural network} (CNN).
The goal of this ISAC-assisted FEEL system is to train\footnote{After the model training stage is completed, each device will deploy the model locally for inference. In this work, we only focus on the optimization in the training stage.} an ML model tailored for specific sensing applications, e.g., human motion recognition, using the data wirelessly sensed by the ISAC devices.

In FEEL, a shared ML model, represented by $\mathbf{w} \in \mathbb{R}^{d}$, is trained collaboratively over all the ISAC devices with the coordination of the edge server. 
Specifically, we seek to find such a model $\mathbf{w}$ that can minimize the objective function, 
\begin{equation}\label{eq:loss}
	F(\mathbf{w}) = \frac{1}{K}\sum_{k\in \mathcal{K}}\mathbb{E}_{\xi \sim \mathcal{P}_{k}}[f_{k}(\mathbf{w};\xi)],
\end{equation}
where $f_{k}(\mathbf{w};\xi)$ is the local loss function at device $k$; $\xi$ is a random seed whose realization represents a batch of samples.

The FEEL training process proceeds iteratively over multiple communication rounds.
In an arbitrary round $r$, mainly five steps are executed as elaborated as follows (see Fig. \ref{fig:fig-system-model}):
\begin{enumerate}
	\item \textbf{Global model broadcasting}: Each ISAC device downloads the global FL model $\mathbf{w}^{(r)}$ from the server via the wireless broadcast channel.
	\item \textbf{Data sensing}: \textcolor{black}{Each ISAC device switches to the sensing mode, and transmits dedicated FMCW signals (as shown in Fig. \ref{fig:fig-sensing-signal-format}) for sensing. Then, a batch of sensing data, i.e., $\lbrace (x_{i},y_{i}) \rbrace_{i \in \mathcal{D}_{k}^{(r)}}$, can be attained at each ISAC device by processing its received echo signals as mentioned.}  This batch of data with size $b^{(r)} = \left\vert \mathcal{D}_{k}^{(r)} \right\vert$ will be used for local computation in the following Step 3. The batch size $b^{(r)}$ can vary adaptively over different rounds, but keep unchanged within any particular round\footnote{The size of batch can also vary adaptively among different devices in each communication round to reduce the waiting time \cite{Ma2021TMC-batch}. In this work, we focus on adjusting batch size over different communication rounds by designing the sensing strategy to accelerate the convergence speed of FEEL.}.
	\item \textbf{Local model updating for $\tau$ steps}: Each device updates its model by running $\tau$ steps of stochastic gradient decent (SGD) from $\mathbf{w}^{(r)}$, i.e.,
	\begin{align*}
		\mathbf{w}_{k}^{(r,0)} &= \mathbf{w}^{(r)},\\
		\mathbf{w}_{k}^{(r,\ell+1)} &= \mathbf{w}_{k}^{(r,\ell)} - \eta \mathbf{g}_{k}(\mathbf{w}_{k}^{(r,\ell)}),\;\ell=0,2,...,\tau-1,\;\forall k,
	\end{align*}
	where $\eta$ is the learning rate; $\mathbf{g}_{k}(\mathbf{w}_{k}^{(r,\ell)}) = \frac{1}{b^{(r)}}\sum_{i \in \mathcal{D}_{k}^{(r)}} \nabla f(\mathbf{w}_{k}^{(r,\ell)};x_{i},y_{i})$ is the stochastic gradient. 
	\item \textbf{Local model uploading}: \textcolor{black}{Each ISAC device switches to communication mode and uploads its local model after $\tau$ local updates, i.e., $\mathbf{w}_{k}^{(r,\tau)}$, to the server via the uplink wireless channel.}
	\item \textbf{Global aggregation}: Once the server receives the models from all the devices, it aggregates them to obtain a new global ML model, i.e., 
	\begin{equation*}
		\mathbf{w}^{(r+1)} = \frac{1}{K}\sum_{k\in\mathcal{K}}\mathbf{w}_{k}^{(r,\tau)}.
	\end{equation*}
\end{enumerate}

%


\subsection{Communication Model}\label{sec:comm-model}

We assume that each device occupies a non-overlapping communication frequency subcarrier, and the transmissions of the devices are interference-free from each other. The received baseband signal from device $k$ at time $t$ is given by
\begin{equation*}
	y_{{\sf c},k}(t) = \sqrt{p_{{\sf c},k}(t)}h_{k}(t)c_{k}(t) + n_{k}(t),
\end{equation*}
where $h_{k}(t) \in \mathbb{C}$ denotes the channel from device $k$ to the server, $p_{{\sf c},k}(t)$ is the communication transmit power, $c_{k}(t)$ is the communication transmit signal, and $n_{k}(t)$ is additive Gaussian white noise. Since the latency for transmitting a model to the server is typically much longer than the channel coherence time, we assume that the communication channels from the devices to the server are fast Rayleigh fading channel \cite{Tse2005Fundamentals}. Specifically, the channel propagation coefficient between the server and device $k$ is generally modeled as\footnote{The time index $t$ will be omitted for simplicity hereafter, unless necessary.} $h_{k} = \sqrt{\phi_{k}}\overline{h}_{k}$; here, $\phi_{k}$ describes the large-scale propagation effects, including path loss and shadowing effect, and $\overline{h}_{k}$ represents the small-scale fading modeled as \textit{independent and identically distributed} (i.i.d.) \textit{circularly symmetric complex Gaussian} (CSCG) random variables with zero mean and unit variance, i.e., $\overline{h}_{k} \sim \mathcal{CN}(0,1)$. We assume that the large-scale propagation coefficient $\phi_{k}$ remains unchanged during the whole uplink transmission, while the small-scale fading $\overline{h}_{k}$ varies from one coherence time block to another. Moreover, we assume that the channel coefficients are only known at the server by channel estimation. The ergodic capacity, instead of instantaneous capacity, of device $k$, is used to quantify the capacity of the fast Rayleigh fading channel between device $k$ and the server \cite{Tse2005Fundamentals}, which is represented as a function of $p_{{\sf c},k}$ :
\begin{equation}\label{eq:ergodic-rate}
	C_{k}(p_{{\sf c},k}) = \mathbb{E}_{h_{k}}\left[B_{0}\log_{2}\left( 1 + \frac{p_{{\sf c},k}\vert h_{k}\vert^{2}}{B_{0}N_{0}} \right)\right],
\end{equation}
where $B_{0}$ is the bandwidth of each subcarrier, and $N_{0}$ is the noise power spectral density. 

\begin{figure*}[!t]
	\centering
	\includegraphics[width=0.65\textwidth]{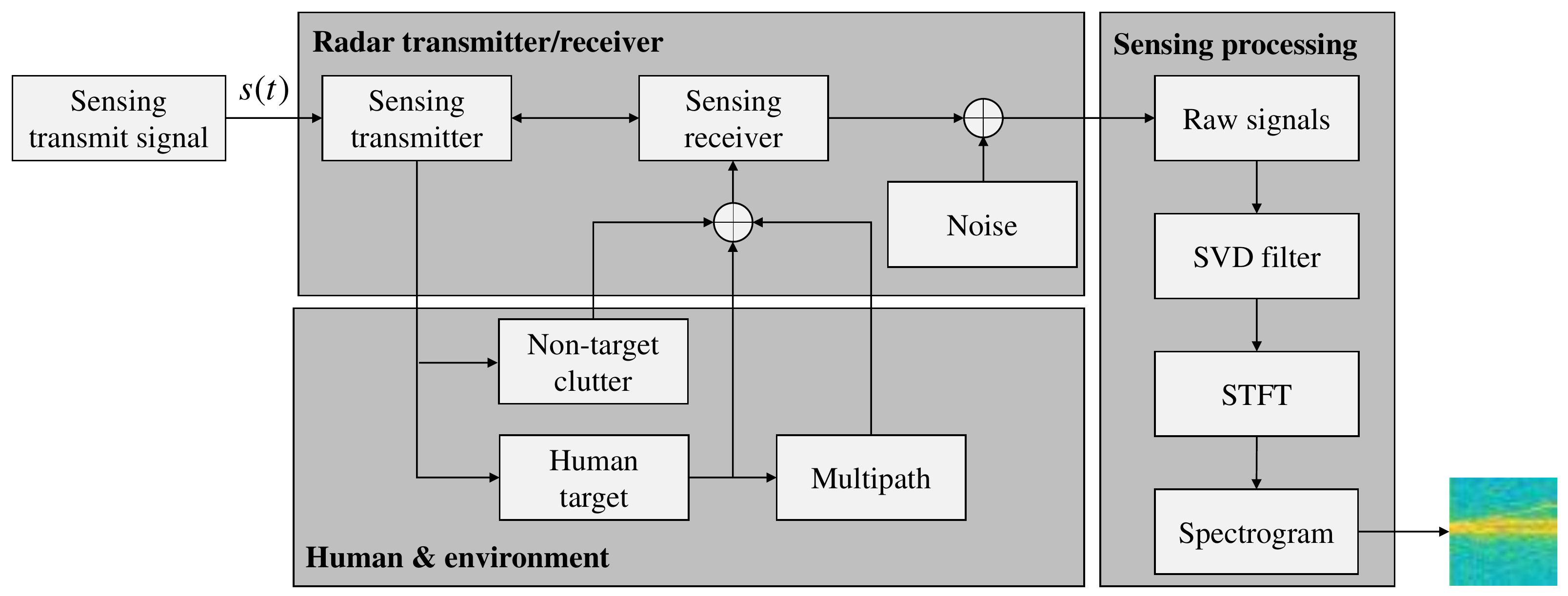}
	\vspace{-0.3cm}
	\caption{The diagram of wireless sensing for human motion recognition.}
	\label{fig:fig-sensing-processing}
\end{figure*}

\newcounter{MYtempeqncnt}
\begin{figure*}[!b]
	\hrulefill
	\normalsize
	\setcounter{MYtempeqncnt}{\value{equation}}
	\setcounter{equation}{2}
	\begin{equation}\label{eq:reflected-signal}
		y_{{\sf s},k,p}(t) =\sqrt{p_{{\sf s},k}(t)}\cdot \underbrace{\frac{A}{\sqrt{4\pi}}\sum_{b=1}^{B}\frac{\sqrt{G_{k,b,p}(t)}}{r_{k,b,p}^{2}(t)}\exp(-j\frac{4\pi f_{\sf c}}{c}r_{k,b,p}(t))s_{k}\left(t-\frac{2r_{k,b,p}(t)}{c}\right)}_{\triangleq \bar{y}_{k,p}(t)},
	\end{equation}
	\setcounter{equation}{\value{equation}}
\end{figure*}



\subsection{Sensing Model}\label{sec:sensing-model}


The FMCW signal for sensing of device $k$ is denoted as $s_{k}(t)$.
The bandwidth of sensing signal is denoted as $B_{\sf s}$, and the duration of each chirp is $T_{\sf p}$, as shown in Fig. \ref{fig:fig-sensing-signal-format}. 
Moreover, we focus on sensing/recognizing dynamic targets such as moving humans in this work. 
We consider the scenario that the devices are geometrically separated at different locations with sufficient distance between each other, such that all the devices can sense the environment over the same bandwidth and will not be interfered with each other\footnote{This is well justified by the fact that the range of the sensing applications such as human/object recognition is usually small, typically less than 20 meters \cite{Cui2021Netw_ISAC}.}.

We use a primitive-based method \cite{Ram2010TGRS} to model the scattering from the entire human body to the device sensing receiver as a linear time varying system. The scattering along $p$-th path is approximated using superposition of the returns from $B$ body primitives as given by (\ref{eq:reflected-signal}) at the bottom of next page,
where $p_{{\sf s},k}(t)$ is the sensing transmit power of device $k$ at time $t$, $A$ is a constant related to the gain of the antenna, $G_{k,b,p}(t)$ is the complex amplitude proportional to the \textit{radar-cross sections} (RCS) of the $b$-th primitive at time $t$ along the $p$-th path, $r_{k,b,p}(t)$ is the distance from the $b$-th primitive to the sensing receiver along the $p$-th path, $c=3 \times 10^{8}$ $m/s$ is the light speed, and $\bar{y}_{k,p}(t)$ is the scattering along $p$-th path normalized by the square root of the sensing transmit power.

Then, the received signal from the environment at the sensing receiver of device $k$ is given by
\begin{equation}\label{eq:sensing-receive-signal}
\begin{aligned}
	 y_{{\sf s},k}(t) &= \sum_{p=1}^{P}\sqrt{p_{{\sf s},k}(t)}\overline{y}_{{\sf s},k,p}(t) + e_{k}(t),\\
	 &= \sqrt{p_{{\sf s},k}(t)}\overline{y}_{{\sf s},k,1}(t) + \sqrt{p_{{\sf s},k}(t)}\overline{y}_{{\sf s},k,-1}(t) + e_{k}(t),
\end{aligned}
\end{equation}
where $\overline{y}_{{\sf s},k,1}(t)$ is the normalized scattering along first-order (direct) reflected path, and $\overline{y}_{{\sf s},k,-1}(t) \triangleq \sum_{p=2}^{P}\overline{y}_{{\sf s},k,p}(t)$ is summation of the normalized scattering along higher-order (indirect) reflected paths, and $e_{k}(t)$ is additive term because of the ground clutter and receiver noise. We assume that the additive ground clutter and receiver noise components are zero-mean complex Gaussian distributed \cite{Sen2014SJ-OFDMsensing}. In practice, the ground clutter is small enough to be ignored in a space where no other dynamic objects exist except one moving target.

\section{Communication and Sensing Analysis}\label{sec:analysis}

In this section, we analyze the ergodic communication rate and obtain its closed-form expression. Moreover, we also look into the wireless sensing process in human motion recognition and analyze the sensing quality.

\subsection{Communication Rate Analysis}

To get a closed-form expression of $C_{k}(p_{{\sf c},k})$ in (\ref{eq:ergodic-rate}), it is rewritten as
\begin{align*}
	\nonumber C_{k}(p_{{\sf c},k}) &= \int_{0}^{+\infty}B_{0}\log_{2}\left(1+\frac{p_{{\sf c},k}x}{B_{0}N_{0}}\right)f_{\vert h_{k} \vert^{2}}(x)dx\\
	&=\frac{B_{0}}{\ln 2}\frac{p_{{\sf c},k}}{B_{0}N_{0}}\int_{0}^{+\infty}\frac{1-F_{\vert h_{k} \vert^{2}}(x)}{1+\frac{p_{{\sf c},k}x}{B_{0}N_{0}}}dx,
\end{align*}
where $f_{\vert h_{k} \vert^{2}}(x)$ and $F_{\vert h_{k} \vert^{2}}(x)$ are the probability density function and cumulative distribution function of the random variable $\vert h_{k} \vert^{2}$, respectively. It can be verified that $\vert h_{k} \vert^{2}$ follows an exponential distribution, i.e., $\vert h_{k} \vert^{2} \sim \text{Exp}(1/\phi_{k})$. Hence, we have $F_{\vert h_{k} \vert^{2}}(x) = 1 - e^{-x/\phi_{k}}$. Then, $C_{k}(p_{{\sf c},k})$ can be calculated as 
\begin{align*}
	C_{k}(p_{{\sf c},k}) &= \frac{B_{0}}{\ln 2}\frac{p_{{\sf c},k}}{B_{0}N_{0}}\int_{0}^{+\infty}\frac{e^{-x/\phi_{k}}}{1+\frac{p_{{\sf c},k}x}{B_{0}N_{0}}}dx\\
	&=\frac{B_{0}}{\ln 2}\int_{0}^{+\infty}\frac{e^{-x/\phi_{k}}}{x+\frac{B_{0}N_{0}}{p_{{\sf c},k}}}dx.
\end{align*}
According to \cite[Section 8.212]{Gradshteyn2014Integrals}, for real number $a$ and $b>0$, $\int_{0}^{+\infty}\frac{e^{-bx}}{a+x}dx=-e^{ab}\text{Ei}(-ab)$, where $\text{Ei}(x)=\int_{-\infty}^{x}\frac{e^{\rho}}{\rho}d\rho$ is the exponential integral function. Then, $C_{k}(p_{{\sf c},k})$ can be rewritten in closed-form as follows:
\begin{equation}\label{eq:rate_close_form}
	C_{k}(p_{{\sf c},k}) = -\frac{B_{0}}{\ln 2}e^{\frac{B_{0}N_{0}}{p_{{\sf c},k}\phi_{k}}}\text{Ei}(-\frac{B_{0}N_{0}}{p_{{\sf c},k}\phi_{k}}).
\end{equation}
It can be verified that $C_{k}(p_{{\sf c},k})$ is an increasing function of $p_{{\sf c},k}$.

\subsection{Sensing Quality Analysis}\label{sec:sensing-quality}

In this subsection, we analyze the quality of sensing samples for human motion recognition.
Micro-Doppler signature is a characteristic of motion, which can be used for recognizing human motions.
A common technique for micro-Doppler analysis is the time-frequency representation, such as spectrograms \cite{Chen2002Book-TF}.
Thus, the received sensing signal as in (\ref{eq:sensing-receive-signal}) needs to be further preprocessed in order to obtain the spectrograms as the sensing data samples. Each spectrogram is generated from a long sequence of the received signal spanning a duration of $T_{0}$, called unit sensing time, which consists of $M$ chirps, as shown in Fig. \ref{fig:fig-sensing-signal-format}. 
We assume that the sensing transmit power will remain unchanged during the duration. The sampling rate is denoted as $f_{\sf s}$. The received raw signal as shown in (\ref{eq:sensing-receive-signal}) after sampling in $i$-th duration can be reconstructed as a 2D sensing data matrix, i.e., $\mathbf{Y}_{{\sf s},k}[i] \in \mathbb{C}^{f_{\sf s}T_{\sf p} \times M}$, which is the superposition of first-order scattering, higher-order scattering, and the clutter and receiver noise in 2D format as follows:
\begin{equation*}
	\mathbf{Y}_{{\sf s},k}[i] = \sqrt{p_{{\sf s},k}[i]}\mathbf{Y}_{{\sf s},k,1}[i] + \sqrt{p_{{\sf s},k}[i]}\mathbf{Y}_{{\sf s},k,-1}[i] + \mathbf{E}_{k}[i],
\end{equation*}
where $[\mathbf{Y}_{{\sf s},k}[i]]_{\ell,m} = y_{{\sf s},k}((i-1)T_{0} + \ell+mT)$, $[\mathbf{Y}_{{\sf s},k,1}[i]]_{\ell,m} = \bar{y}_{{\sf s},k,1}((i-1)T_{0} + \ell+mT)$, $[\mathbf{Y}_{{\sf s},k,-1}[i]]_{\ell,m}=\bar{y}_{{\sf s},k,-1}((i-1)T_{0} + \ell+mT)$, $[\mathbf{E}_{k}[i]]_{\ell,m} = e_{k}((i-1)T_{0} + \ell+mT)$, $\ell \in [0,f_{\sf s}T_{\sf p}-1]$ represents the fast time index, and $m \in [0,M-1]$ represents the slow time index. 


As shown in Fig. \ref{fig:fig-sensing-processing}, similar to \cite{Li2021SPAWC-sensing-model}, we extract the useful first-order scattering\footnote{The higher-order scattering can be utilized as a diverse source to improve the sensing performance \cite{Zhang2019multipath}, which, however, makes the processing more complicated. In this work, the high-order scattering signals are treated as interference to the first-order scattering when generating the spectrograms.} from $\mathbf{Y}_{{\sf s},k}[i]$ via \textit{singular value decomposition} (SVD) based filter. Suppose that the SVD of $\mathbf{Y}_{{\sf s},k}$ is denoted as\footnote{The index for each duration is omitted for simplicity, since the signal at each duration will be preprocessed in the same way.} $\mathbf{Y}_{{\sf s},k} = \mathbf{U}_{k}\mathbf{\Sigma}_{k}\mathbf{V}_{k}^{\ast}$ where $\mathbf{U}_{k} \in \mathbb{C}^{f_{\sf s}T_{\sf p} \times f_{\sf s}T_{\sf p}}$ and $\mathbf{V}_{k} \in \mathbb{C}^{M \times M}$ are complex unitary matrices, and $\mathbf{\Sigma}_{k} \in \mathbb{C}^{f_{\sf s}T_{\sf p} \times M}$ is  rectangular diagonal matrix with singular values on the diagonal. The signal after filtering is $\mathbf{Y}_{{\sf F},k} = [\mathbf{U}_{k}]_{:,r_{a}:r_{b}}[\mathbf{\Sigma}_{k}]_{r_{a}:r_{b},:}\mathbf{V}_{k}^{\ast}$, where $r_{a}$ and $r_{b}$ are pre-determined hyperparameters. It is equivalent to write $\mathbf{Y}_{{\sf F},k}$ as
\begin{align*}
	\mathbf{Y}_{{\sf F},k} &= \tilde{\mathbf{U}}_{k}\mathbf{Y}_{{\sf s},k}\tilde{\mathbf{V}}_{k}^{\ast}\\
	& = \sqrt{p_{{\sf s},k}}\tilde{\mathbf{U}}_{k}\mathbf{Y}_{{\sf s},k,1}\tilde{\mathbf{V}}_{k}^{\ast} + \sqrt{p_{{\sf s},k}}\tilde{\mathbf{U}}_{k}\mathbf{Y}_{{\sf s},k,-1}\tilde{\mathbf{V}}_{k}^{\ast} + \tilde{\mathbf{U}}_{k}\mathbf{E}_{k}\tilde{\mathbf{V}}_{k}^{\ast},
\end{align*}
where $\tilde{\mathbf{U}}_{k} = \left([\mathbf{U}_{k}]_{:,r_{a}:r_{b}}\right)\left([\mathbf{U}_{k}]_{:,r_{a}:r_{b}}\right)^{\ast}$ and $\tilde{\mathbf{V}}_{k} = \left([\mathbf{V}_{k}]_{:,r_{a}:r_{b}}\right)\left([\mathbf{V}_{k}]_{:,r_{a}:r_{b}}\right)^{\ast}$. Denote $\tilde{\mathbf{Y}}_{{\sf s},k,1} = \tilde{\mathbf{U}}_{k}\mathbf{Y}_{{\sf s},k,1}\tilde{\mathbf{V}}_{k}^{\ast}$, $\tilde{\mathbf{Y}}_{{\sf s},k,-1} = \tilde{\mathbf{U}}_{k}\mathbf{Y}_{{\sf s},k,-1}\tilde{\mathbf{V}}_{k}^{\ast}$, and $\tilde{\mathbf{E}}_{k} = \tilde{\mathbf{U}}_{k}\mathbf{E}\tilde{\mathbf{V}}_{k}^{\ast}$. Then, $\mathbf{Y}_{{\sf F},k}$ is rewritten as 
\begin{equation*}
	\mathbf{Y}_{{\sf F},k} = \sqrt{p_{{\sf s},k}}\tilde{\mathbf{Y}}_{{\sf s},k,1} + \sqrt{p_{{\sf s},k}}\tilde{\mathbf{Y}}_{{\sf s},k,-1} + \tilde{\mathbf{E}}_{k}.
\end{equation*}

Next, \textit{short-time Fourier transform} (STFT) with a sliding window function $\omega[n]$ of length $W$ is adopted to generate the \textit{range-Doppler-time} (RDT) cube in order to obtain time-frequency features of $\mathbf{Y}_{{\sf F},k}$. The RDT cube of $\mathbf{Y}_{{\sf F},k}$ after STFT is given by
\begin{align*}
	S_{k}[\ell,f,m] = \sum_{n=0}^{W}[\mathbf{Y}_{{\sf F},k}]_{\ell,m(W-Q)-n}\exp(-j2\pi f n/W)\omega[n],
\end{align*}
where $Q$ is the number of overlapping points; $f \in [0,W-1]$ and $m \in \left[1,\frac{M-Q}{W-Q}\right]$ are the frequency and temporal shift indexes, respectively. Then, we non-coherently integrate $S_{k}[\ell,f,m]$ within all the range bins and obtain the integrated STFT of $\mathbf{Y}_{{\sf F},k}$ as
\begin{equation*}
	\overline{S}_{k}[f,m] = \sum_{\ell=0}^{f_{\sf s}T}S_{k}[\ell,f,m].
\end{equation*}
Due to the linearity of STFT and summation operation, $\overline{S}_{k}[f,m]$ can be given by
\begin{equation}\label{eq:STFT-expression}
	\overline{S}_{k}[f,m] = \sqrt{p_{{\sf s},k}}\overline{S}_{k,1}[f,m] + \sqrt{p_{{\sf s},k}}\overline{S}_{k,-1}[f,m] + \overline{S}_{e,k}[f,m],
\end{equation}
where $\overline{S}_{k,1}[f,m]$, $\overline{S}_{k,-1}[f,m]$, and $\overline{S}_{e,k}[f,m]$ are integrated STFT of $\tilde{\mathbf{Y}}_{{\sf s},k,1}$, $\tilde{\mathbf{Y}}_{{\sf s},k,-1}$, and $\tilde{\mathbf{E}}_{k}$, respectively. The first-order scattering related term $\sqrt{p_{{\sf s},k}}\overline{S}_{k,1}[f,m]$ represents useful information in the spectrogram; the higher-order scattering related term $\sqrt{p_{{\sf s},k}}\overline{S}_{k,-1}[f,m]$ and the clutter and noise related term $\overline{S}_{e,k}[f,m]$ corrupt the spectrogram as interference.


\begin{figure}[!t]
	\centering
	\includegraphics[width=0.32\textwidth]{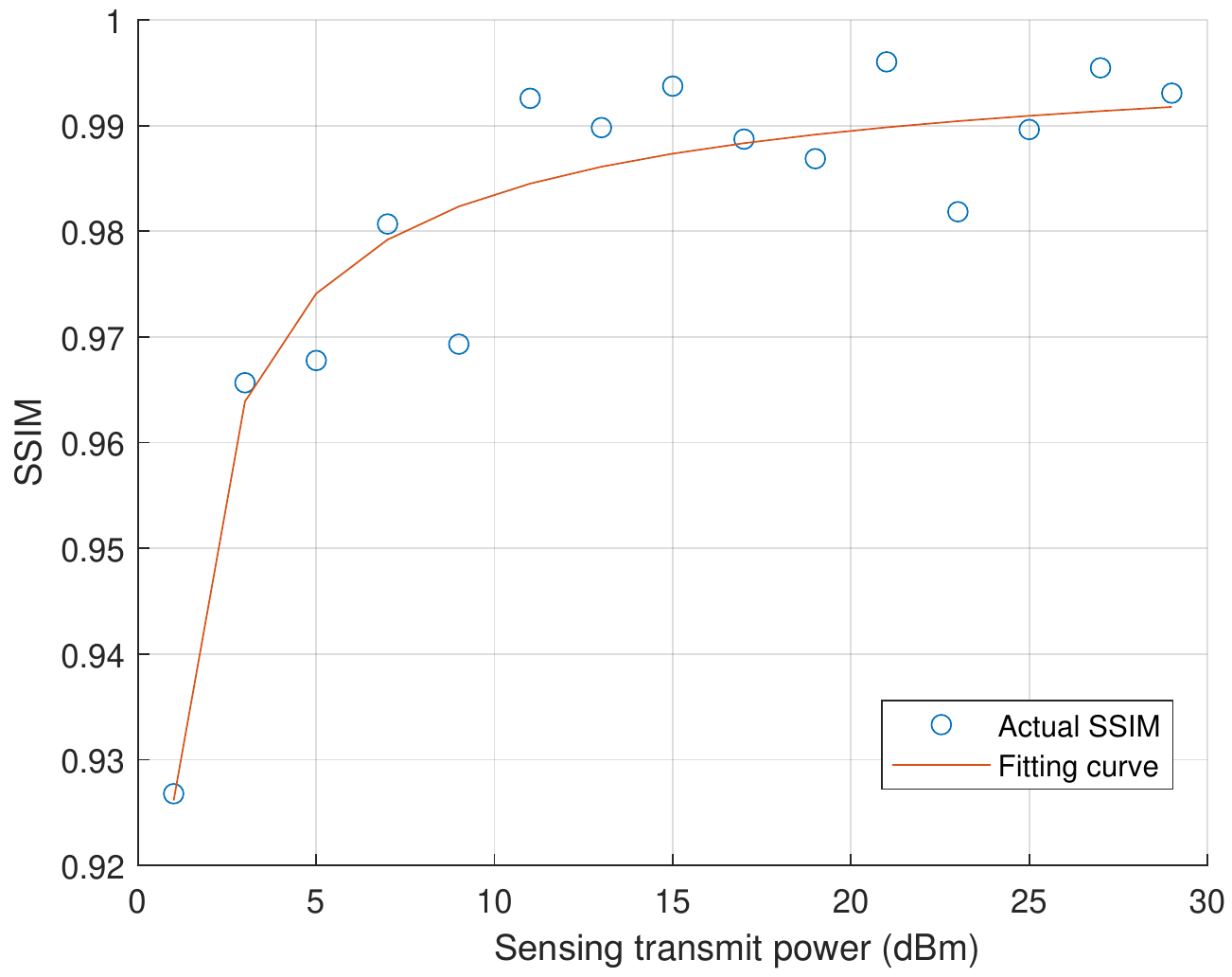}
	\caption{The spectrogram quality (SSIM) versus sensing transmit power.}\vspace{-0.5cm}
	\label{fig:ssim}
\end{figure}

\begin{remark}[Impact of sensing transmit power to the quality of spectrograms]\label{remark:sensing-power}
	From (\ref{eq:STFT-expression}), we can see that the quality of spectrograms will improve as we increase the sensing transmit power. Moreover, when the sensing transmit power is large enough such that the effects of clutter and noise is ignorable compared with that of high-order scattering, the quality of spectrogram will no longer improve. This is also validated by the experimental result in Fig. \ref{fig:ssim}. We use the \textit{structural similarity} (SSIM) index\footnote{SSIM is a metric that measures the perceptual difference between two similar images, which is widely used in image compression and computer vision. It is a full reference metric that requires two images from the same image capture, i.e., a reference image and a processed image.} to visualize the quality of spectrogram \cite{Wang2004TIP-ssim}. Therefore, we can set a threshold for transmit power, say $P_{{\sf s},k}^{\min}$, and if we choose the transmit power as $p_{{\sf s},k} \ge P_{{\sf s},k}^{\min}$, each device can generate data samples (spectrograms) with approximately the same satisfactory quality.
\end{remark}

\section{Problem Formulation}\label{sec:problem-formulation}

In this paper, we aim at accelerating the training process, i.e., minimizing the global loss function $F(\mathbf{w})$ in (\ref{eq:loss}), under the constraints on latency, energy, and peak power. In the following, these three kinds of constraints are formulated.

\subsection{Latency Constraints}

\begin{figure*}[!t]
	\centering
	\includegraphics[width=0.75\textwidth]{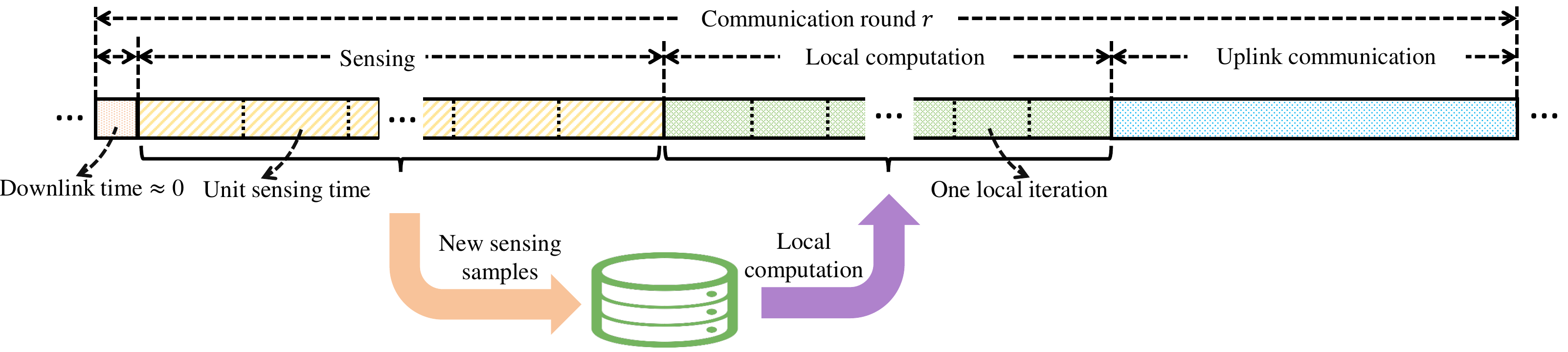}
	\caption{Time allocation within one particular communication round for each ISAC device.}
	\label{fig:fig-model-communication-round}\vspace{-0.5cm}
\end{figure*}

The latency for each device in a communication round comprises four parts: the global model downloading time, sensing time, local computation time, and local model uploading time, as shown in Fig. \ref{fig:fig-model-communication-round}. Since the server broadcasts the same global model to all the devices over the entire frequency band, and the server usually has greater transmit power than the devices, we can ignore the global downloading time in a communication round. In the next, we formulate the other three parts.

\subsubsection{Sensing time} 

Recall that the unit sensing time, i.e., the time for generating a sensing sample, is $T_{0}$, and the number of samples generated at communication round $r$ is $b^{(r)}$. Thus, the sensing time of device $k$ at round $r$ is given by
\begin{equation}\label{eq:sensing-time}
	T_{{\sf s},k}^{(r)} = T_{0} \cdot b^{(r)}.
\end{equation}

\subsubsection{Local computation time}

Let $\nu$ denote the CPU cycles that required for executing the computation of a single sample in the local update. Then, the local computation time of device $k$ at communication round $r$ can be expressed as
\begin{equation}\label{eq:computation-time}
	T_{{\sf cp},k}^{(r)} = \frac{b^{(r)}\nu\tau}{f_{{\sf cpu}}},
\end{equation}
where $f_{{\sf cpu}}$ is the CPU frequency (cycles/s) of each device.

\subsubsection{Local model uploading time}

Since the dimensions of the local model are fixed for all devices, the data size in bits that each device needs to be uploaded is a constant, which is denoted by $D_{\sf b}$. Then, the local model uploading time $T_{{\sf cm},k}$ of device $k$ are equal over all the communication rounds and should satisfy that
\begin{equation*}
	\text{(C1)  }T_{{\sf cm},k}C_{k}(p_{{\sf c},k}) \ge D_{\sf b},\; \forall k \in \mathcal{K},
\end{equation*}
where $C_{k}(p_{{\sf c},k})$ is the ergodic capacity in (\ref{eq:ergodic-rate}).

We consider synchronous aggregation at the server, i.e., the aggregation happens until the local model from all the devices are received. Then, the latency for round $r$ is given by
\begin{equation}\label{eq:one-round-time}
	T^{(r)} = \max_{k\in\mathcal{K}}\left\{ T_{{\sf s},k}^{(r)} + T_{{\sf cp},k}^{(r)} + T_{{\sf cm},k} \right\}.
\end{equation}
Accordingly, the total latency is constrained by
\begin{equation*}
	\text{(C2)  }\sum_{r=1}^{R}T^{(r)} \leq T^{\sf max},
\end{equation*}
where $R$ is the total number of communication rounds in the training process, and $T^{\sf max}$ is the total latency constraint.

\subsection{Energy Constraint}

The energy consumption of each device comes from three parts: energy for sensing, energy for local computation, and energy for local model uploading, which are formulated one by one as follows:

\subsubsection{Energy for sensing}

The energy consumption for sensing of device $k$ at round $r$ is 
\begin{equation}\label{eq:energy-sensing}
	E_{{\sf s},k}^{(r)} = T_{{\sf cp},k}^{(r)}p_{{\sf s},k} = T_{0} b^{(r)} p_{{\sf s},k}.
\end{equation}

\subsubsection{Energy for local computation}

According to \cite{Mao2016JSAC-energy}, the energy consumption for local computation is formulated as
\begin{equation}\label{eq:energy-computation}
	E_{{\sf cp},k}^{(r)} = \tau \theta \nu f_{{\sf cpu}}^{2}b^{(r)},
\end{equation} 
where $\theta$ is effective switched capacitance that depends on the chip architecture of the device.

\subsubsection{Energy for local model uploading}

The energy consumption for local model uploading of device $k$ at round $r$ is calculated as
\begin{equation}\label{eq:energy-comm}
	E_{{\sf cm},k}^{(r)} = T_{{\sf cm},k}p_{{\sf c},k}.
\end{equation}

From practical consideration, the ISAC devices have limited battery life, and the total energy consumption to complete a training task should be under constrained, i.e.,
\begin{equation*}
	\text{(C3)  }\sum_{r=1}^{R}\left(E_{{\sf s},k}^{(r)} + E_{{\sf cp},k}^{(r)} + E_{{\sf cm},k}^{(r)}\right) \leq E^{\sf max},\; \forall k \in \mathcal{K},
\end{equation*}
where $E^{\sf max}$ is the total energy constraint.

\subsection{Peak Power Constraints}

Limited by the transmitter, the communication transmit power should be constrained by 
\begin{equation*}
	\text{(C4)  }0 \leq p_{{\sf c},k} \leq P_{{\sf c},k}^{\max},\; \forall k \in \mathcal{K},
\end{equation*}
where $P_{{\sf c},k}^{\max}$ is the peak communication transmit power.
Moreover, according to Remark \ref{remark:sensing-power}, the sensing transmit power should be constrained by 
\begin{equation*}
	\text{(C5)  }P_{{\sf s},k}^{\min} \leq p_{{\sf s},k} \leq P_{{\sf s},k}^{\max},\; \forall k \in \mathcal{K},
\end{equation*}
where $P_{{\sf s},k}^{\max}$ is the peak sensing transmit power.

\subsection{Problem Formulation}

Under the three kinds of constraints above, the problem of accelerating the training process can be formulated as
\begin{align}
 \nonumber(\text{P1})\;\;\;\underset{\{b^{(r)}\},\{p_{{\sf c},k}\},\{p_{{\sf s},k}\},\{T_{{\sf cm},k}\}}{\text{min   }}
	\;\;\;&\underset{r\in \{1,2,\ldots,R\}}{\text{min   }}F(\mathbf{w}^{(r)}) \\  \text{s.t.} \;\;\;
	&\nonumber \text{(C1)}-\text{(C5)}.
\end{align}

It is challenging to solve Problem (P1) due to: (i) The loss function $F(\mathbf{w}^{(r)})$ at communication round $r\in \mathcal{R} \triangleq \{1,2,\ldots,R\}$ has no exact analytical expression; (ii) Generally, we need to find an expression of learning error with adaptive batch size over $R$ communication rounds in order to solve Problem (P1), which is also a non-trivial task; (iii) The constraints (C2) and (C3) are non-convex. Moreover, although $C_{k}(p_{{\sf c},k})$ has a closed-form expression in (\ref{eq:rate_close_form}), it involves special function which is non-trivial for analysis. In this work, we will derive an upper bound of the learning error for tractability, instead of the exact loss function in Problem (P1), which is beneficial for jointly allocating SC$^{2}$ resource.

\section{Joint SC$^{2}$ Resource Allocation}\label{sec:solution}

In this section, we first present an error upper bound with fixed batch size, based on which we split the original Problem (P1) into two separate subproblems, one optimizing the sensing and communication transmit power and the communication time to maximize the total sensed data samples, and the other optimizing the batch sizes for each round under given maximized total sensed data samples.

\subsection{Error Upper Bound Analysis}

To facilitate the analysis, the following commonly adopted assumptions on local loss functions are made \cite{Wang2018CoopSGD,Wang2018AdaBat,Nori2021TC-fast,Bottou2018Optimization}.

\begin{assumption}[Smoothness]\label{asp:smoothness}
	The global loss function $F(\mathbf{w})$'s is $L$-smooth: for all $\mathbf{w}_{i}$ and $\mathbf{w}_{j}$, $F(\mathbf{w}_{i}) \leq F(\mathbf{w}_{j}) + \left(\mathbf{w}_{i} - \mathbf{w}_{j}\right)^{T}\nabla F(\mathbf{w}_{j}) + \frac{L}{2}\left\Vert \mathbf{w}_{i} - \mathbf{w}_{j} \right\Vert^{2}$.
\end{assumption}

\begin{assumption}[Unbiased gradient with bounded variance]\label{asp:gradient}
	Stochastic gradient $\mathbf{g}_{k}(\mathbf{w})$ of each ISAC device is unbiased with $\sigma^{2}$-uniformly bounded variance in $\ell_{2}$ norm, namely
	\begin{align*}
		&\mathbb{E}[\mathbf{g}_{k}(\mathbf{w})] = \nabla F(\mathbf{w}),\\
		&\mathbb{E}[\left\Vert \mathbf{g}_{k}(\mathbf{w}) - \nabla F(\mathbf{w})\right\Vert^{2}] \leq \frac{\beta}{b}\left\Vert \nabla F(\mathbf{w})\right\Vert^{2} + \frac{\sigma^{2}}{b},
	\end{align*}
	where $b$ is the batch size in calculating gradient $\mathbf{g}_{k}(\mathbf{w})$.
\end{assumption}

\begin{assumption}[Lower bounded]\label{asp:lower_bounded}
	The global loss function is lower bounded by $F_{\sf inf}$, i.e., $F(\mathbf{w}) \ge F_{\sf inf}$, $\forall \mathbf{w}$.
\end{assumption}

Under the above assumptions, the following theorem gives an upper bound on the average-squared gradient norm of FL with fixed batch size.


\begin{lemma}[Error upper bound with fixed batch size \cite{Wang2018CoopSGD}]\label{lemma:error-upper-bound}
	Under Assumptions \ref{asp:smoothness}-\ref{asp:lower_bounded}, with local updates $\tau$ and batch size $b$, if the learning rate satisfies $\eta L + \eta^{2} L^{2} \tau(\tau - 1) \leq 1$, the local ML models are all initialized at the same point $\mathbf{w}_{1}$, then after $R$ communication rounds, the average-squared gradient norm will be bounded by:
	\begin{align}
		\nonumber&\mathbb{E}\left[\frac{1}{\tau R} \sum_{r=1}^{R} \sum_{\ell=1}^{\tau}\Vert \nabla F(\bar{\mathbf{w}}^{(r,l)})\Vert^{2}\right] \\
		\leq & \label{eq:error-upper-bound} \frac{2\left(F(\mathbf{w}^{(1)}) - F_{\sf inf}\right)}{\eta R \tau} + \frac{\eta L \sigma^{2}}{Kb}+\frac{\eta^{2}L^{2}\sigma^{2}\tau}{b},
	\end{align}
where $\bar{\mathbf{w}}^{(r,l)}=\frac{1}{K}\sum_{k\in\mathcal{K}}\mathbf{w}_{k}^{(r,\ell)}$.
\end{lemma}
\begin{IEEEproof}
See Appendix of Theorem 1 in \cite{Wang2018CoopSGD}.
\end{IEEEproof}

\begin{remark}
	To accelerate the training convergence, one needs to minimize the error upper bound in Lemma \ref{lemma:error-upper-bound}. Obviously, increasing the batch size can accelerate the convergence. The explanation is straightforward that larger batch size leads to smaller stochastic gradient noise, which makes the training converge faster \cite{Bottou2018Optimization}.
\end{remark}

In this work, we consider adaptive batch size $b^{(r)}$ over different communication rounds, instead of fixed batch size.
Thus, Lemma \ref{lemma:error-upper-bound} cannot be directly applied to solve Problem (P1). Nevertheless, Lemma \ref{lemma:error-upper-bound} is helpful for splitting Problem (P1) as shown in the following subsection.

\subsection{Problem Splitting}

This subsection shows that Problem (P1) can be split into two subproblems equivalently to make Problem (P1) tractable.

First, substituting (\ref{eq:sensing-time}), (\ref{eq:computation-time}), and (\ref{eq:one-round-time}) into (C2), it follows that
\begin{equation}\label{eq:C2-sub}
	\sum_{r=1}^{R}b^{(r)}(T_{0} + \frac{\nu\tau}{f_{\sf cpu}}) + R \max_{k\in\mathcal{K}}T_{{\sf cm},k} \leq T^{\sf max}.
\end{equation}
After proper manipulation, we can rewrite (\ref{eq:C2-sub}) as
\begin{equation}\label{eq:C2-re}
	b_{\sf sum}\left(T_{0} + \frac{\nu\tau}{f_{\sf cpu}}\right) + R T_{{\sf cm},k} \leq T^{\sf max},\; \forall k \in \mathcal{K},
\end{equation}
where $b_{\sf sum} = \sum_{r=1}^{R}b^{(r)}$. Similarly, we substitute (\ref{eq:energy-sensing}), (\ref{eq:energy-computation}), and (\ref{eq:energy-comm}) into (C3) and can have that 
\begin{equation}\label{eq:C3-re}
	b_{\sf sum}\left(T_{0} p_{{\sf s},k} + \tau \theta \nu f_{{\sf cpu}}^{2}\right) + RT_{{\sf cm},k}p_{{\sf c},k} \leq E^{\sf max},\; \forall k \in \mathcal{K}.
\end{equation}
Therefore, the original Problem (P1) can be rewritten as
\begin{align}
	\nonumber(\text{P2})\;\;\;&\underset{\substack{\{b^{(r)}\},b_{\sf sum},\{p_{{\sf c},k}\},\\ \{p_{{\sf s},k}\},\{T_{{\sf cm},k}\}}}{\text{min   }}
	\;\;\;\underset{r\in \{1,2,\ldots,R\}}{\text{min   }}F(\mathbf{w}^{(r)}) \\  \text{s.t.} \;\;\;
	&\nonumber b_{\sf sum} = \sum_{r=1}^{R}b^{(r)},\\ &\nonumber\text{(C1)},\;\text{(C4)},\;\text{(C5)},\;(\ref{eq:C2-re}),\;\text{and}\;(\ref{eq:C3-re}).
\end{align}

Before we split Problem (P2), the following corollary is given to unveil how $b_{\sf sum}$ can affect the convergence speed.

\begin{corollary}\label{coro}
	Under the given total latency and energy constraint, larger $b_{\sf sum}$ leads to smaller average-squared gradient norm, i.e., faster convergence speed.
\end{corollary}

\begin{IEEEproof}
	From Lemma \ref{lemma:error-upper-bound}, the average-squared gradient norm after one communication round, say the $r$-th communication round, is upper bounded by
	\begin{align*}
		&\mathbb{E}\left[\frac{1}{\tau } \sum_{\ell=1}^{\tau}\Vert \nabla F(\bar{\mathbf{w}}^{(r,l)})\Vert^{2}\right]\\
		\leq & \frac{2\left(F(\mathbf{w}^{(r)}) - F_{\sf inf}\right)}{\eta \tau} + \frac{\eta L \sigma^{2}}{Kb^{(r)}}+\frac{\eta^{2}L^{2}\sigma^{2}\tau}{b^{(r)}}.
	\end{align*}
	It follows that larger batch size $b^{(r)}$ leads to smaller average-squared gradient norm after the training of the $r$-th communication round. Assume that the optimal batch size of the $r$-th communication round is $b^{\ast(r)}$. Obviously, if there exists $\hat{b}_{\sf sum} > b_{\sf sum}$, one can always find a set of batch sizes $\{\hat{b}^{(r)}\}$ such that $\sum_{r=1}^{R}\hat{b}^{(r)}=\hat{b}_{\sf sum}$ and $\hat{b}^{(r)} \ge b^{\ast(r)}$ for any $r \in \mathcal{R}$. Therefore, compared with $b_{\sf sum}$, $\hat{b}_{\sf sum}$ can lead to smaller average-squared gradient norm, i.e., faster convergence speed, which completes the proof.
\end{IEEEproof}

Intuitively, $b_{\sf sum}$ is the summation of the batch sizes at all the communication rounds for each device, and also the number of data samples generated/sensed by each device in the training process. From Corollary \ref{coro}, the devices needs to generate as many data samples as possible to accelerate the convergence speed. However, on the other hand, generating more data samples means greater consumption of the time and energy resources. Since the devices are time and energy constrained, the total number of generated data samples is limited.

Obviously, $b_{\sf sum}$ only depends on the constraints in Problem (P2). From Corollary \ref{coro}, we need to find maximum $b_{\sf sum}$ under the constraints in Problem (P2) in order to maximize the convergence speed. Therefore, Problem (P2) can be equivalently solved by: (i) first optimizing $\{p_{{\sf c},k}\}$, $\{p_{{\sf s},k}\}$, and $\{T_{{\sf cm},k}\}$ to maximize $b_{\sf sum}$ under the constraints in Problem (P2); (ii) then optimizing $\{b^{(r)}\}$ given maximum $b_{\sf sum}$ to maximize the convergence speed.

\subsection{Maximization of Total Sensed Data Samples}

In the following, we solve the first subproblem above to maximize the total sensed data samples, i.e., $b_{\sf sum}$, subject to all the constraints in Problem (P2), which is formulated as 
\begin{align}
	\nonumber(\text{P3})\;\;\;&\underset{b_{\sf sum},\{p_{{\sf c},k}\},\{p_{{\sf s},k}\},\{T_{{\sf cm},k}\}}{\text{max   }}
	\;\;\;b_{\sf sum} \\  \text{s.t.} \;\;\;
	&\nonumber \text{(C1)},\;\text{(C4)},\;\text{(C5)},\;(\ref{eq:C2-re}),\;\text{and}\;(\ref{eq:C3-re}).
\end{align}
Problem (P3) is challenging to solve due to: (i) Constraints (C1), (\ref{eq:C2-re}), and (\ref{eq:C3-re}) are non-convex; (ii) Constraints (C1), (\ref{eq:C2-re}), and (\ref{eq:C3-re}) involves $C_{k}(p_{{\sf c},k})$ in (\ref{eq:rate_close_form}), which has complex special function. It is generally impossible to solve Problem (P3) by using standard techniques for convex optimization. In the following, we first give the optimal sensing transmit power $\{p_{{\sf s},k}\}$, and then simplify Problem (P3) such that we can use grid search efficiently to find the optimal solutions.


From Constraint (\ref{eq:C3-re}) we have that
\begin{equation*}
	b_{\sf sum} \leq \frac{E^{\sf max} - RT_{{\sf cm},k}p_{{\sf c},k}}{T_{0} p_{{\sf s},k} + \tau \theta \nu f_{{\sf cpu}}^{2}} ,\; \forall k \in \mathcal{K},
\end{equation*}
which gives an upper bound of $b_{\sf sum}$. Moreover, this upper bound is decreasing with $p_{{\sf s},k}$. Combined with Constraint (C5), we can obtain that the optimal sensing transmit power $p_{{\sf s},k}^{\ast}$ in Problem (P3) is $p_{{\sf s},k}^{\ast} = P_{{\sf s},k}^{\min}$.

Then, Constraint (\ref{eq:C3-re}) in Problem (P3) becomes
\begin{equation}\label{eq:C3-re-re}
	b_{\sf sum} \leq \frac{E^{\sf max} - RT_{{\sf cm},k}p_{{\sf c},k}}{T_{0} P_{{\sf s},k}^{\min} + \tau \theta \nu f_{{\sf cpu}}^{2}} ,\; \forall k \in \mathcal{K}.
\end{equation}
Constraint (\ref{eq:C2-re}) can also be transformed into
\begin{equation}\label{eq:C2-re-re}
	b_{\sf sum} \leq \frac{T^{\sf max} - R T_{{\sf cm},k}}{T_{0} + \frac{\nu\tau}{f_{\sf cpu}}},\; \forall k \in \mathcal{K}.
\end{equation}
Then, (\ref{eq:C3-re-re}) and (\ref{eq:C2-re-re}) can be combined as
\begin{equation}\label{eq:b-sum-combine}
	b_{\sf sum} \leq \min_{k \in \mathcal{K}}\min \left\lbrace \frac{T^{\sf max}-RT_{{\sf cm},k}}{T_{0} + \frac{\nu\tau}{f_{\sf cpu}}},\frac{E^{\sf max}-RT_{{\sf cm},k}p_{{\sf c},k}}{T_{0} P_{{\sf s},k}^{\min} + \tau \theta \nu f_{{\sf cpu}}^{2}} \right\rbrace.
\end{equation}

\begin{proposition}\label{prop2}
	Constraint (C1) in Problem (P3) is equivalent to
	\begin{equation}\label{eq:pc-equal}
		T_{{\sf cm},k}C_{k}(p_{{\sf c},k}) = D_{\sf b},\; \forall k \in \mathcal{K}.
	\end{equation}
\end{proposition}
\begin{IEEEproof}
	If the equality in (C1) is not satisfied, i.e., $T_{{\sf cm},k}C_{k}(p_{{\sf c},k}) > D_{\sf b}$, it follows from (\ref{eq:b-sum-combine}) that one can always decease $T_{{\sf cm},k}$ and thus increase $b_{\sf sum}$.
\end{IEEEproof}

Then, substituting (\ref{eq:pc-equal}) into (\ref{eq:b-sum-combine}) yields
\begin{equation}
	b_{\sf sum} \leq \min_{k \in \mathcal{K}}\Phi_{k}(p_{{\sf c},k}),
\end{equation}
where
\begin{equation}\label{eq:phi-k}
	\Phi_{k}(p_{{\sf c},k}) \triangleq \min \left\lbrace \frac{T^{\sf max}-\frac{RD_{\sf b}}{C_{k}(p_{{\sf c},k})}}{T_{0} + \frac{\nu\tau}{f_{\sf cpu}}},\frac{E^{\sf max}-\frac{Rp_{{\sf c},k}D_{\sf b}}{C_{k}(p_{{\sf c},k})}}{T_{0} P_{{\sf s},k}^{\min} + \tau \theta \nu f_{{\sf cpu}}^{2}} \right\rbrace,
\end{equation}
which is a function of $p_{{\sf c},k}$.
With the above results, Problem (P3) reduces to a single variable optimization problem regarding to $\{p_{{\sf c},k}\}$ as follows
\begin{align}
	\nonumber(\text{P4})\;\;\;\underset{\{p_{{\sf c},k}\}}{\text{max   }}
	\;\;\;& \min_{k \in \mathcal{K}}\Phi_{k}(p_{{\sf c},k}) \\  \text{s.t.} \;\;\;
	&\nonumber \text{(C4)}.
\end{align}
Since $\{p_{{\sf c},k}\}$ are independent with each other for $k \in \mathcal{K}$, it can be verified that the optimal communication transmit power is given by
\begin{equation}\label{eq:pc-optimal}
p_{{\sf c},k}^{\ast} = \arg\max_{p_{{\sf c},k}} \Phi_{k}(p_{{\sf c},k}),\; \forall k \in \mathcal{K}.
\end{equation}
It is still challenging to obtain $p_{{\sf c},k}^{\ast}$ from (\ref{eq:pc-optimal}) directly, since $\Phi_{k}(p_{{\sf c},k})$ is still complicated. In this work, we tend to find $p_{{\sf c},k}^{\ast}$ by grid search in $\left[0,P_{{\sf c},k}^{\max}\right]$. After we have $\{p_{{\sf c},k}\}$, the optimal $b_{\sf sum}$ is obtained by substituting $\{p_{{\sf c},k}^{\ast}\}$ into (\ref{eq:b-sum-combine}), i.e., 
\begin{equation}\label{eq:optimal-b-sum}
	b_{\sf sum} = \min_{k \in \mathcal{K}}\Phi_{k}(p_{{\sf c},k}^{\ast}).
\end{equation}
It follows from (\ref{eq:pc-equal}) that the optimal communication time is given by
\begin{equation}\label{eq:optimal-time}
	T_{{\sf cm},k}^{\ast} = \frac{D_{\sf b}}{C_{k}(p_{{\sf c},k}^{\ast})},\; \forall k \in \mathcal{K}.
\end{equation}

\begin{remark}[Energy-limited and latency-limited scenarios]\label{remark:limited}
	When the energy constraint $E^{\sf max}$ is large enough, the FEEL system will be latency-limited, and the latency constraint will be tight, i.e., $\Phi_{k}(p_{{\sf c},k})$ in (\ref{eq:phi-k}) reduces to $\Phi_{k}(p_{{\sf c},k}) = \min \left\lbrace \frac{T^{\sf max}-RD_{\sf b}/C_{k}(p_{{\sf c},k})}{T_{0} + \nu\tau/f_{\sf cpu}} \right\rbrace$. Since $C_{k}(p_{{\sf c},k})$ is an increasing function of $p_{{\sf c},k}$, $\Phi_{k}(p_{{\sf c},k})$ also increases with $p_{{\sf c},k}$. Thus, the optimal communication transmit power in this case is $p_{{\sf c},k}^{\ast} = P_{{\sf c},k}^{\max}$. Alternatively, when the latency constraint $T^{\sf max}$ is large enough, the FEEL system will be energy-limited, and the energy constraint will be tight, and $\Phi_{k}(p_{{\sf c},k})$ in (\ref{eq:phi-k}) reduces to $\Phi_{k}(p_{{\sf c},k}) \triangleq \min \left\lbrace \frac{E^{\sf max}-Rp_{{\sf c},k}D_{\sf b}/C_{k}(p_{{\sf c},k})}{T_{0} P_{{\sf s},k}^{\min} + \tau \theta \nu f_{{\sf cpu}}^{2}} \right\rbrace$. However, it is difficult to verify the monotonicity of $\Phi_{k}(p_{{\sf c},k})$ in this case, and we can only find optimal $p_{{\sf c},k}^{\ast}$ by grid search.
\end{remark}

\subsection{Adaptive Batch Size Optimization}

After obtaining the maximum number of total sensed data samples, i.e., $b_{\sf sum}^{\ast}$, we tend to solve the second subproblem, which optimizes $\{b^{(r)}\}$ under given $b_{\sf sum}^{\ast}$ to maximize the convergence speed.

For given $b_{\sf sum}^{\ast}$, Problem (P2) becomes
\begin{align}
	\nonumber(\text{P5})\;\;\;\underset{\{b^{(r)}\}}{\text{min   }}
	\;\;\;&\underset{r\in \{1,2,\ldots,R\}}{\text{min   }}F(\mathbf{w}^{(r)}) \\  \text{s.t.} \;\;\;
	&\nonumber \sum_{r=1}^{R}b^{(r)} = b_{\sf sum}^{\ast}.
\end{align}
Problem (P5) is equivalent to allocating batch size over different communication rounds given the total sensed data samples. It is generally not possible to obtain the exact expression of the loss function at each round, not to mention that the analytical relation between the loss function and the batch size at each round. The common method is to replace the loss function with an error upper bound; however, we need to find a expression of learning error that can be represented by an function of adaptive batch size over $R$ communication rounds, which is also non-trivial as mentioned earlier. Thus, Problem (P5) cannot be solved directly. Instead, in this work, we propose an alternative approach to approximately optimize the adaptive batch size at each communication round based on the following proposition.

\begin{proposition}[Error upper bound for given total sensed samples] \label{prop:2}
	Under Assumptions \ref{asp:smoothness}-\ref{asp:lower_bounded}, with local updates $\tau$ and batch size $b$, if the learning rate satisfies $\eta L + \eta^{2} L^{2} \tau(\tau - 1) \leq 1$, the local FL models are all initialized at the same point $\mathbf{w}_{1}$, then after $R^{\prime}$ communication rounds, or sensing $b_{\sf sum}^{\prime}=b R^{\prime}$ samples at each device, the minimal expected squared gradient norm will be bounded by:
	\begin{align}
		\nonumber&\mathbb{E}\left[\frac{1}{\tau R} \sum_{r=1}^{R} \sum_{\ell=1}^{\tau}\Vert \nabla F(\bar{\mathbf{w}}^{(r,l)})\Vert^{2}\right]\\
		\label{eq:error-upper-bound-given-total-batch}\leq & \frac{2\left(F(\mathbf{w}^{(1)}) - F_{\sf inf}\right)b}{\eta \tau b_{\sf sum}^{\prime}} + \frac{\eta L \sigma^{2}}{Kb}+\frac{\eta^{2}L^{2}\sigma^{2}\tau}{b},
	\end{align}
	where $\bar{\mathbf{w}}^{(r,l)}=\frac{1}{K}\sum_{k\in\mathcal{K}}\mathbf{w}_{k}^{(r,\ell)}$.
\end{proposition}
\begin{IEEEproof}
	It follows from $b_{\sf sum}^{\prime}=b R^{\prime}$ that $R^{\prime} = \frac{b_{\sf sum}^{\prime}}{b}$. Substitute it into (\ref{eq:error-upper-bound}) in Lemma \ref{lemma:error-upper-bound}, and we complete the proof.
\end{IEEEproof}

It can be verified that, to minimize the error upper bound in Proposition \ref{prop:2}, i.e., the right-hand side of (\ref{eq:error-upper-bound-given-total-batch}), the optimal batch size $b$ is given by
\begin{align}\label{eq:optimal-batch}
	b^{\ast} = \sqrt{\frac{\eta^{2} L  \sigma^{2} \tau b_{\sf sum}^{\prime}\left(1+\eta K L\tau\right)}{2K\left(F(\mathbf{w}^{(1)}) - F_{\sf inf}\right)}}.
\end{align}

The result in (\ref{eq:optimal-batch}) can be applied in the design of adaptive batch size to accelerate the convergence speed. According to (\ref{eq:optimal-batch}), the best choice of $b^{(1)}$ in the $1$-st communication round should be 
\begin{align}
	\label{eq:optimal-batch-step-t0} b^{(1)} = \sqrt{\frac{\eta^{2} L  \sigma^{2} \tau b_{\sf sum}^{\prime}\left(1+\eta K L\tau\right)}{2K\left(F(\mathbf{w}^{(1)}) - F_{\sf inf}\right)}}.
\end{align}
For the $r$-th communication round, it can be viewed that each device restarts the training at a new initial point $\mathbf{w}^{(r)}$. Similarly, the best choice of $b^{(r)}$ at the $r$-th communication round should be 
\begin{align}
	\label{eq:optimal-batch-step-tl} b^{(r)} = \sqrt{\frac{\eta^{2} L  \sigma^{2} \tau b_{\sf sum}^{\prime}\left(1+\eta K L\tau\right)}{2K\left(F(\mathbf{w}^{(r)}) - F_{\sf inf}\right)}}.
\end{align}
Combining the results in (\ref{eq:optimal-batch-step-t0}) and (\ref{eq:optimal-batch-step-tl}), and approximating $F_{\sf inf}$ by $0$, it yields that\footnote{Since we consider a classification task, the loss value typically approximates $0$ when the training converge, which is also validated by experiments in Sec. \ref{sec:simulation}.}
\begin{align}
	\label{eq:b-tau-update} b^{(r)} \approx \sqrt{\frac{F(\mathbf{w}^{(1)})}{F(\mathbf{w}^{(r)})}}b^{(1)}.
\end{align}

\begin{remark}[Adaptive batch size]
	The results in (\ref{eq:b-tau-update}) simply reveals an updating rule for the batch size from one communication round to the next. That is, the batch size should increase as the value of loss function gets smaller. 
\end{remark}

The loss function value $F(\mathbf{w}^{(r)})$ can be easily obtained in the training process, and the only problem is to determine the initial batch size $b^{(1)}$ at the 1-st communication round such that  $\sum_{r=1}^{R}b^{(r)} = b_{\sf sum}^{\ast}$. One way to get the value of $b^{(1)}$ is to perform one-dimensional grid search over all possible values of $b^{(1)}$, which is inefficient. In the following, we propose a heuristic scheme, but efficient and practical for implementation.
From empirical results, the value of loss function decreases approximately in an order of $\mathcal{O}(1/r)$ when training with SGD \cite{Peter2019advances}. It follows from (\ref{eq:b-tau-update}) that  the optimal batch size $b^{(r)}$ at communication round $r$ should increase in an order of $\mathcal{O}(\sqrt{r})$, and it can be approximated by
\begin{equation}\label{eq:br-approx}
	b^{(r)} \approx \alpha\sqrt{r} + b_{0},
\end{equation}
where $b_{0}$ is a hyperparameter that determines the batch size at the initial communication round, and $\alpha$ is a constant such that $\sum_{r=1}^{R}b^{(r)} = b_{\sf sum}^{\ast}$ holds. Then, we have that
\begin{equation*}
	\sum_{r=1}^{R}\left(\alpha\sqrt{r} + b_{0}\right) \approx b_{\sf sum}^{\ast}.
\end{equation*}
Thus, $\alpha$ is calculated by
\begin{equation}\label{eq:b0-approx}
	\alpha \approx \frac{b_{\sf sum} - b_{0}R}{\sum_{r=1}^{R}\sqrt{r}}.
\end{equation}
Substituting (\ref{eq:b0-approx}) into (\ref{eq:br-approx}), and considering that $b^{(r)}$'s are integers, we can approximate $b^{(r)}$ by
\begin{equation}\label{eq:ada-batch}
	b^{(r)} \approx \left\lfloor \frac{\left(b_{\sf sum} - b_{0}R\right)\sqrt{r}}{\sum_{r=1}^{R}\sqrt{r}} + b_{0}\right\rfloor.
\end{equation}

\subsection{Overall joint SC$^{2}$ resource allocation scheme}\label{sec:overall-method}

In general, the joint SC$^{2}$ resource allocation scheme by solving Problem (P1) consists of two steps as follows:
\begin{itemize}
	\item \textbf{Step 1:} We optimize the sensing and communication transmit power and the communication time to obtain the maximum total sensed samples $b_{\sf sum}^{\ast}$, where the optimal sensing power is given by $p_{\sf{s},k}^{\ast} = P_{\sf{s},k}^{\sf{min}}$, the optimal communication power $p_{\sf{c},k}^{\ast}$ is obtained by solving the problem in (\ref{eq:pc-optimal}), the optimal communication time $T_{{\sf cm},k}^{\ast}$ is obtained from (\ref{eq:optimal-time}), and the maximum total sensed samples $b_{\sf sum}^{\ast}$ is obtained from (\ref{eq:optimal-b-sum});
	\item \textbf{Step 2:} We determine the batch size of each communication round following (\ref{eq:ada-batch}) with $b_{\sf sum}^{\ast}$ from Step 1.
\end{itemize}

\section{Simulation Results}\label{sec:simulation}

In this section, we present the simulation results to validate the proposed joint SC$^2$ resource allocation scheme. The simulation parameters are summarized in Table \ref{tab:sensing-parameters}.

\subsection{Simulation Setup}


\begin{table}[!t]
	\caption{Simulation Parameters }
	\label{tab:sensing-parameters}
	\centering
	\begin{tabular}{l l}
		\hline
		\bfseries Parameter & \bfseries Value\\
		\hline\hline
		Number of ISAC devices, $K$ & $6$ \\
		Maximum communication transmit power, $P_{{\sf c},k}^{\max}$ & $20$ dBm\\
		Minimum sensing transmit power, $P_{{\sf s},k}^{\min}$ & $20$ dBm\\
		Variance of shadow fading, $\sigma_{\zeta}^{2}$ & 8 dB \\
		Noise power spectral density, $N_{0}$ & -174 dBm/Hz \\
		Communication bandwidth for each device, $B_{0}$ & 0.5 MHz\\
		CPU frequency, $f_{\sf cpu}$ & $5 \times 10^{8}$ cycles/s \\
		CPU cycles for one sample, $\nu$ & $2.5 \times 10^{7}$ \\
		Effective switched capacitance of CPU, $\theta$ & $10^{-27}$ \\
		Number of local updates, $\tau$ & 10 \\
		Total communication rounds, $R$ & 300 \\
		Bandwidth for sensing, $B_{\sf s}$ & $10$ MHz \\
		Carrier frequency, $f_{\sf c}$ & $60$ GHz\\
		Chirp duration, $T_{\sf p}$ & $10$$\mu s$ \\
		Chirp numbers per frame, $M$ & 25 \\
		Unit sensing time, $T_{0}$ & $0.5$ s\\
		Sampling rate, $f_{\sf s}$ & $10$ MHz \\
		\hline
	\end{tabular}
\end{table}

\subsubsection{FEEL system} In this experiment, we consider a FEEL network consisting of an edge server and $K=6$ ISAC devices. The devices are randomly located in a circular area of radius $500$ meters. The large-scale propagation coefficient in dB from device $k$ to the server is modeled as $[\phi_{k}]_{\text{dB}}=[\text{PL}_{k}]_{\text{dB}} +  [\zeta_{k}]_{\text{dB}}$, where $[\text{PL}_{k}]_{\text{dB}} = 128.1+37.6\log_{10}\text{dist}_{k}$ ($\text{dist}_{k}$ is the distance in kilometer) is the path loss in dB, and $[\zeta_{k}]_{\text{dB}}$ is the shadow fading in dB \cite{Yang2020TWC-pathloss}. In this simulation, $[\zeta_{k}]_{\text{dB}}$ is Gauss-distributed random variable with mean zero and variance $\sigma^{2}_{\zeta}$. 

\subsubsection{Sensing task} 
\begin{figure}[!t]
	\centering
	\includegraphics[width=0.38\textwidth]{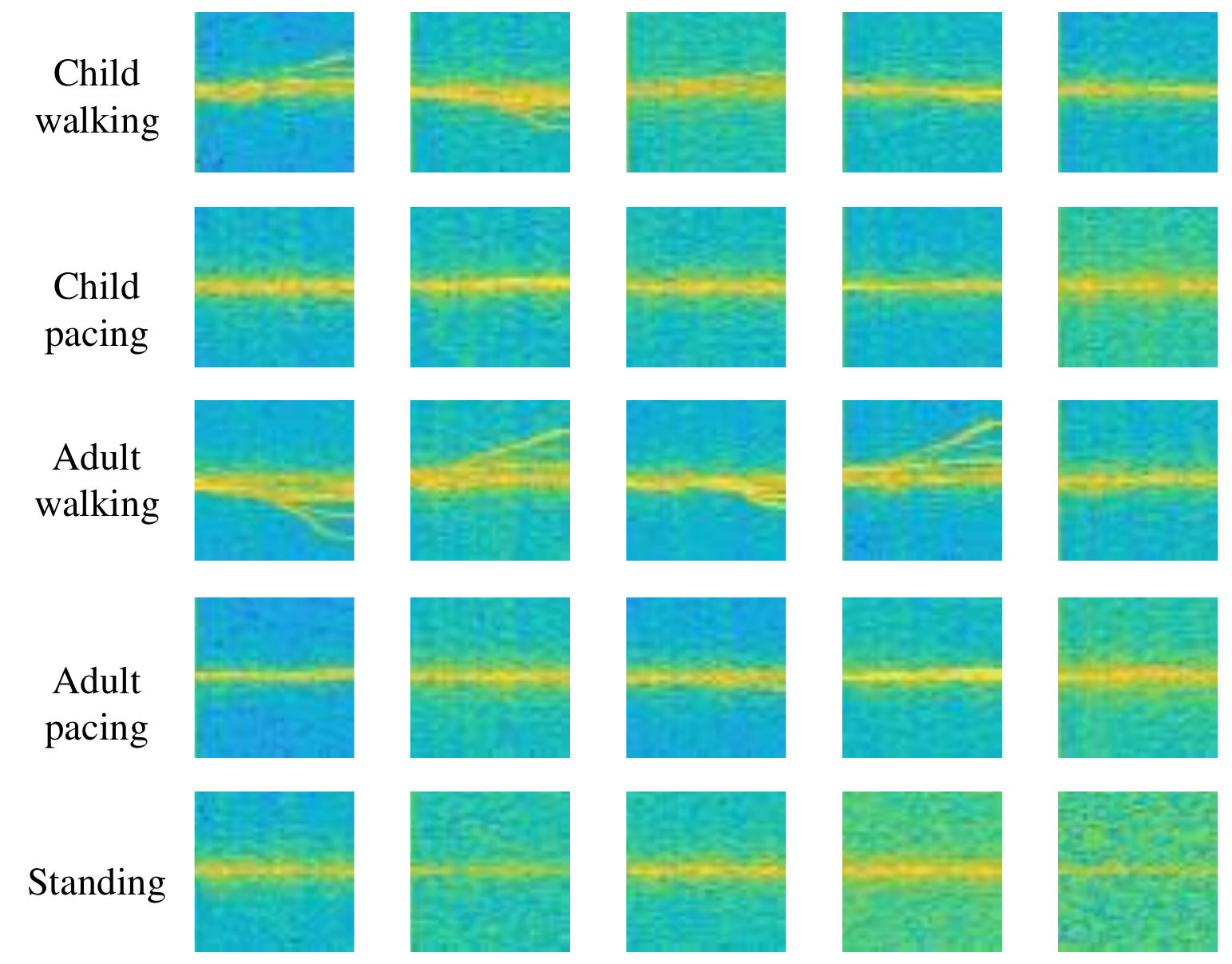}
	\caption{Example data samples of each human motion.}\vspace{-0.3cm}
	\label{fig:fig-data-samples}
\end{figure}

We apply the wireless sensing simulator in \cite{Li2021SPAWC-sensing-model} to simulate various high-fidelity human motions and generate human motion datasets.
In our simulation, the sensing task is to identify five different human motions, i.e., \textit{child walking}, \textit{child pacing}, \textit{adult walking}, \textit{adult pacing}, and \textit{standing}. The heights of children and adults are uniformly distributed in interval $[0.9\text{m}, 1.2\text{m}]$ and $[1.6\text{m}, 1.9\text{m}]$, respectively. The speed of standing, walking, and pacing are $0$ m/s, $0.5H$ m/s, and $0.25H$ m/s, respectively, where $H$ is the height value. The heading of the moving human is set to be uniformly distributed in $[–180^{\circ}, 180^{\circ}]$. Some example data samples of each human motion are shown in Fig. \ref{fig:fig-data-samples}.

\subsubsection{Learning model} We apply widely used ResNet-10 (4,900,677 model parameters in total) with batch normalization as the classifier model \cite{He2016ResNet}. The learning rate is $0.1$.


\subsection{Baselines}

In the simulations, we evaluate our proposed joint SC$^{2}$ resource allocation scheme as summarized in Sec. \ref{sec:overall-method} by comparing with the following baseline schemes:

\textbf{Baseline 1}: (I-BS-MaxPower) This scheme naively uses maximum communication transmit power $P_{{\sf c},k}^{\max}$ in Step 1, and the obtained total sensing samples is denoted as $b_{\sf sum}^{\prime}$. However, increasing batch size of $b^{(r)}=\left\lfloor \frac{\left(b_{\sf sum}^{\prime} - b_{0}R\right)\sqrt{r}}{\sum_{r=1}^{R}\sqrt{r}} + b_{0}\right\rfloor$ is considered in Step 2 as same as (\ref{eq:ada-batch}) in our proposed scheme.  By comparing with this baseline, we evaluate the validity of Step 1 in our proposed scheme.

\textbf{Baseline 2}: (E-BS-OptimalPower) This schemes optimizes each transmit power in Step 1 as same as in our proposed scheme, but considers equal batch size in Step 2. Specifically, the batch size at communication round $r$ is calculated as $b^{(r)} =\left\lfloor \frac{b_{\sf sum}^{\ast}}{R} \right\rfloor$. By comparing with this baseline, we evaluate the validity of Step 2 in our proposed scheme.

\textbf{Baseline 3}: (D-BS-OptimalPower) The only difference with Baseline 2 is that this scheme considers decreasing batch size instead in Step 2. Specifically, the batch size at communication round $r$ is calculated as $b^{(r)} = \left\lfloor \frac{\left(b_{\sf sum}^{\ast} - b_{0}R\right)\sqrt{R-r+1}}{\sum_{r=1}^{R}\sqrt{r}} + b_{0}\right\rfloor$. By comparing with this baseline, we also evaluate the validity of Step 2 in our proposed scheme.

\subsection{Simulation Results}

%

\begin{figure}[!t]
	\centering
	\includegraphics[width=0.36\textwidth]{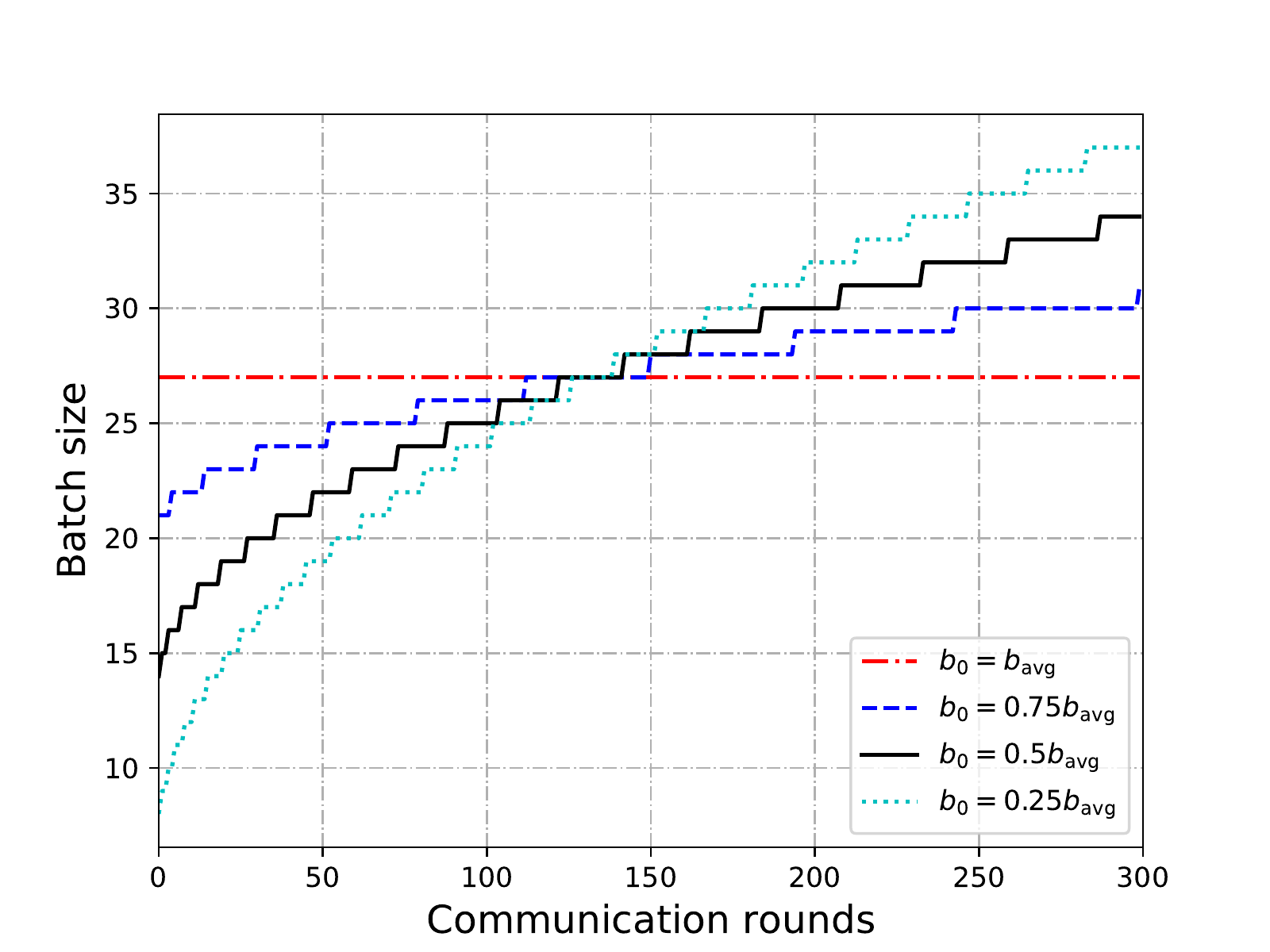}
	\caption{The batch size at each communication round under different values of hyperparameter $b_{0}$.}
	\label{fig:fig_batch_div}
\end{figure}

\begin{figure}[!t]
	\centering
	\includegraphics[width=0.36\textwidth]{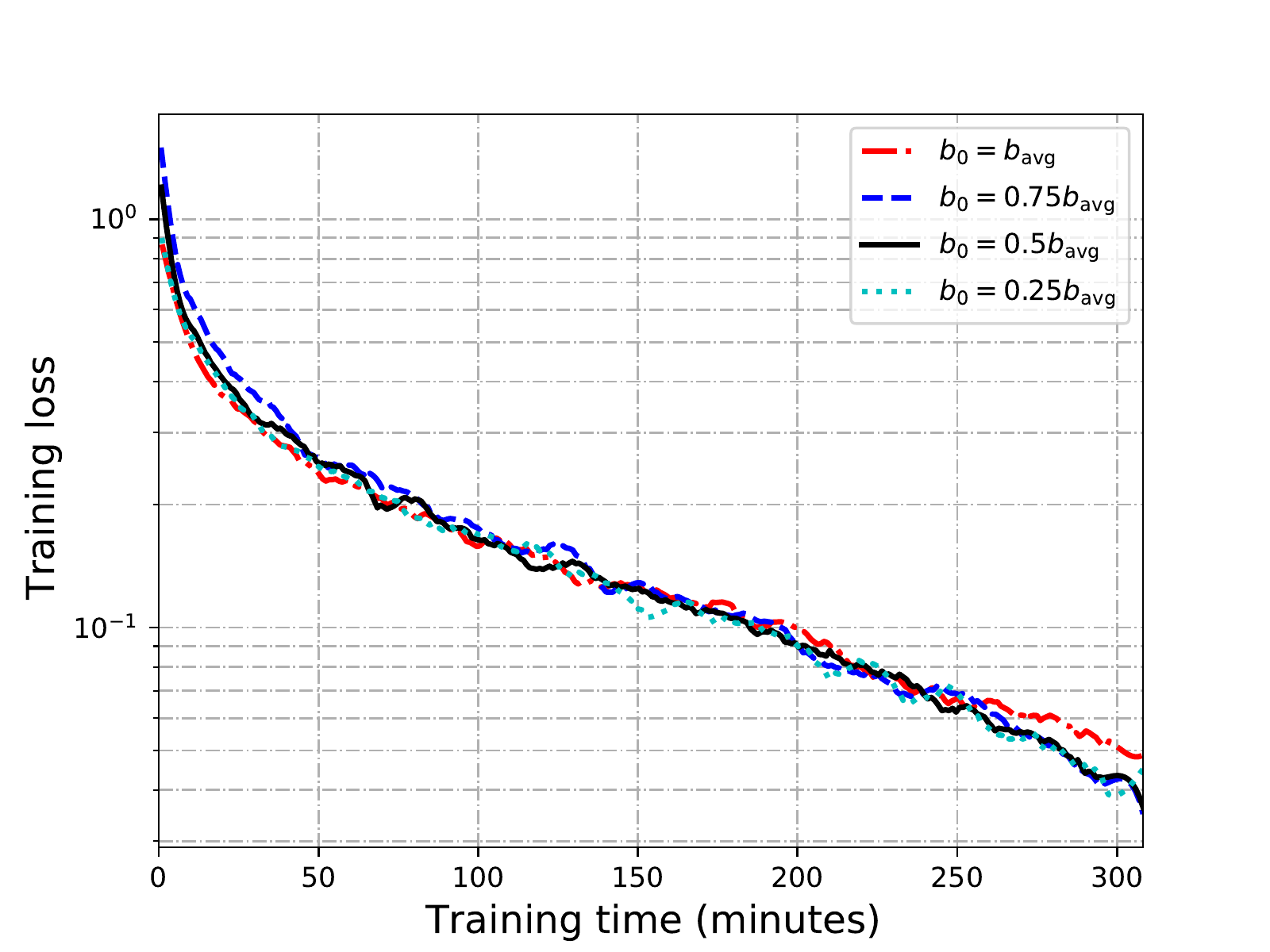}
	\caption{Training loss versus training time under different values of hyperparameter $b_{0}$.}
	\label{fig:fig_training_loss_batch}
\end{figure}

\begin{figure}[!t]
	\centering
	\includegraphics[width=0.36\textwidth]{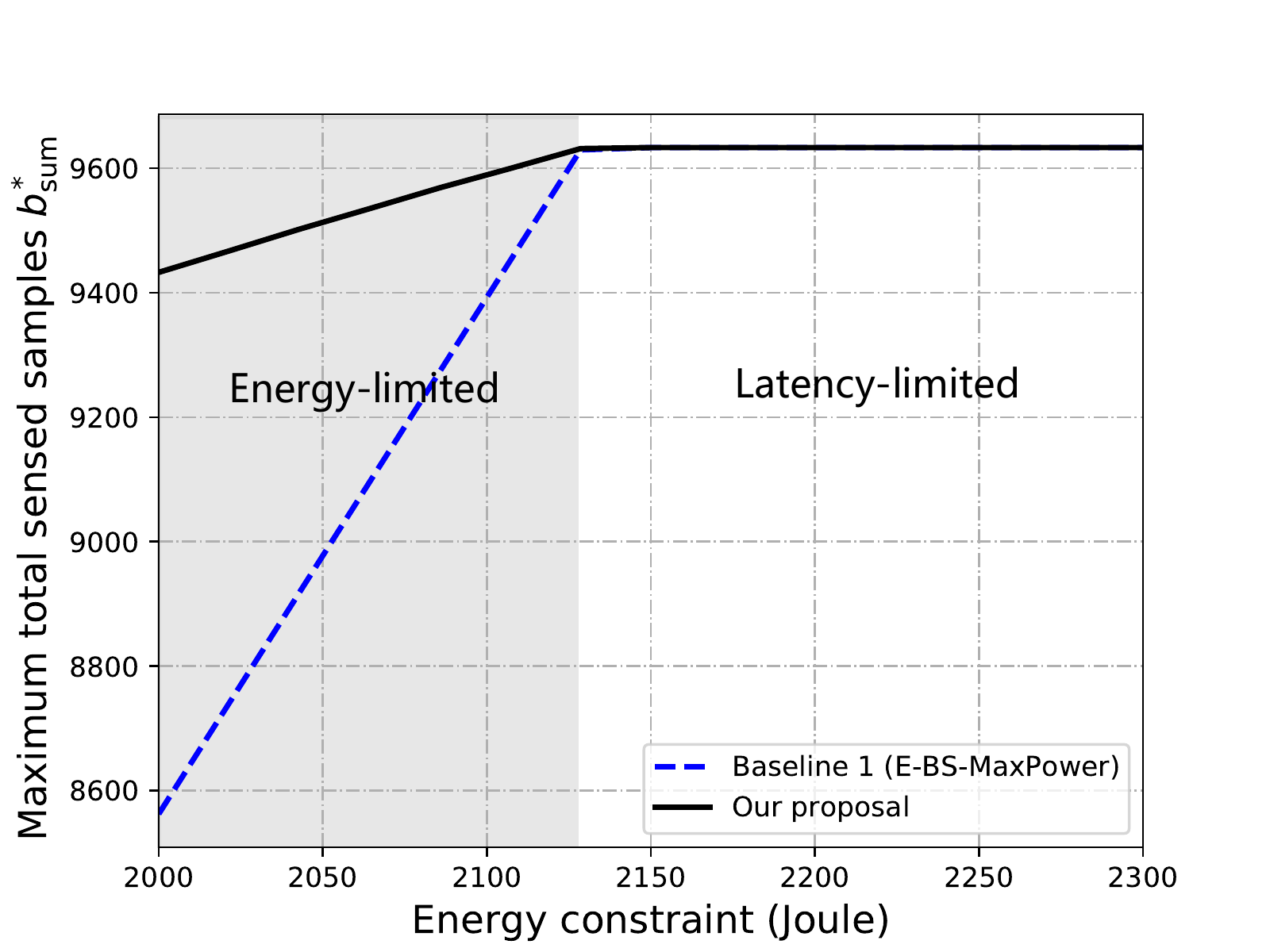}
	\caption{Maximum total sensed samples $b_{\sf sum}^{\ast}$ versus energy constraints under fixed latency constraint $T^{\sf max}=20000$ seconds.}
	\label{fig:fig_bsum_energy}
\end{figure}

\begin{figure}[!t]
	\centering
	\includegraphics[width=0.36\textwidth]{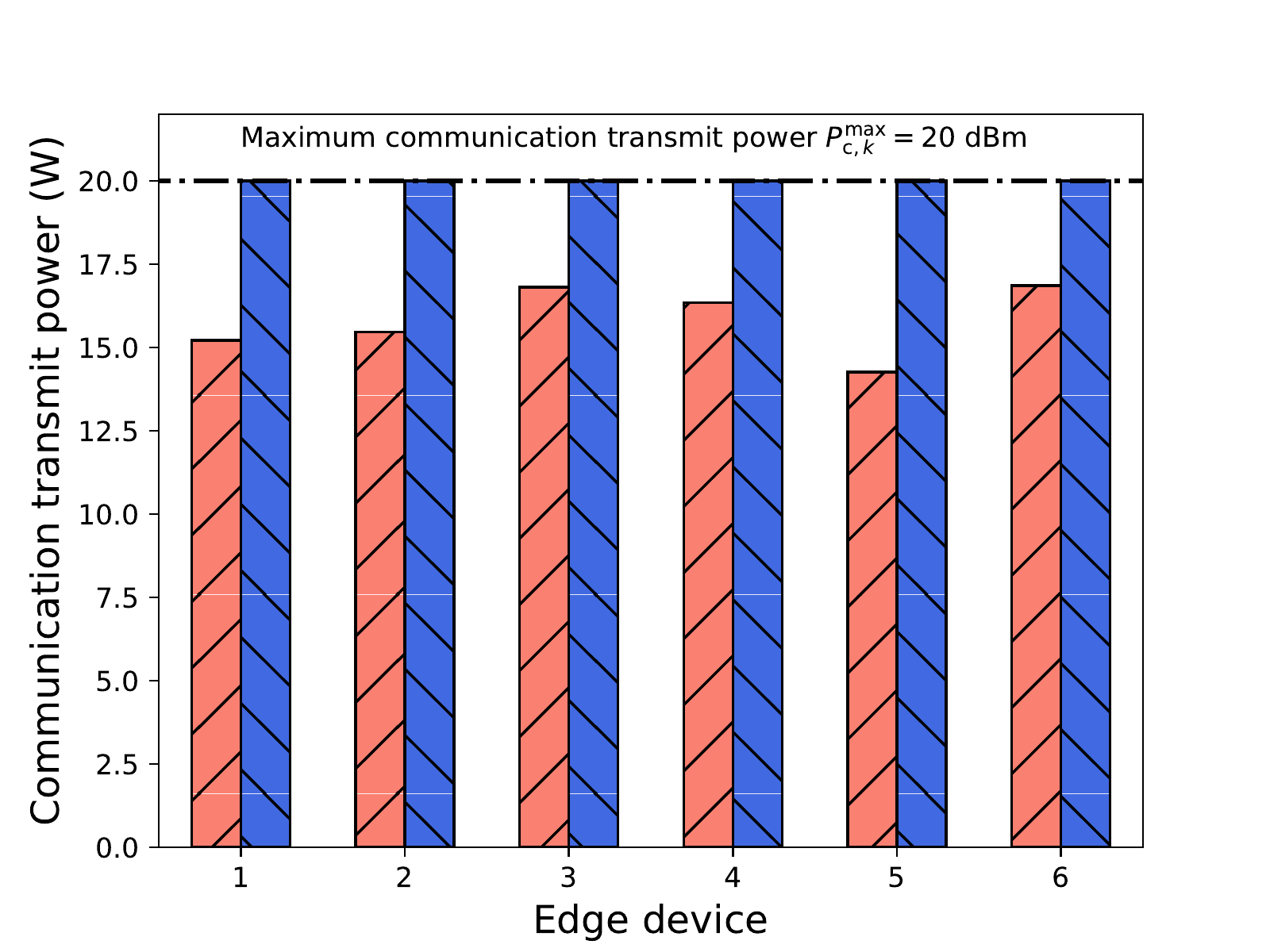}
	\caption{Optimal communication transmit power of each device. (Left: energy-limited case, $\left(T^{\sf max}, E^{\sf max}\right)=\left(20000\text{ s}, 1500\text{ J}\right)$; right: latency-limited case, $\left(T^{\sf max}, E^{\sf max}\right)=\left(20000\text{ s}, 2200\text{ J}\right)$.}
	\label{fig:fig_power_EL}
\end{figure}

\begin{figure}[!t]
	\centering
	\includegraphics[width=0.36\textwidth]{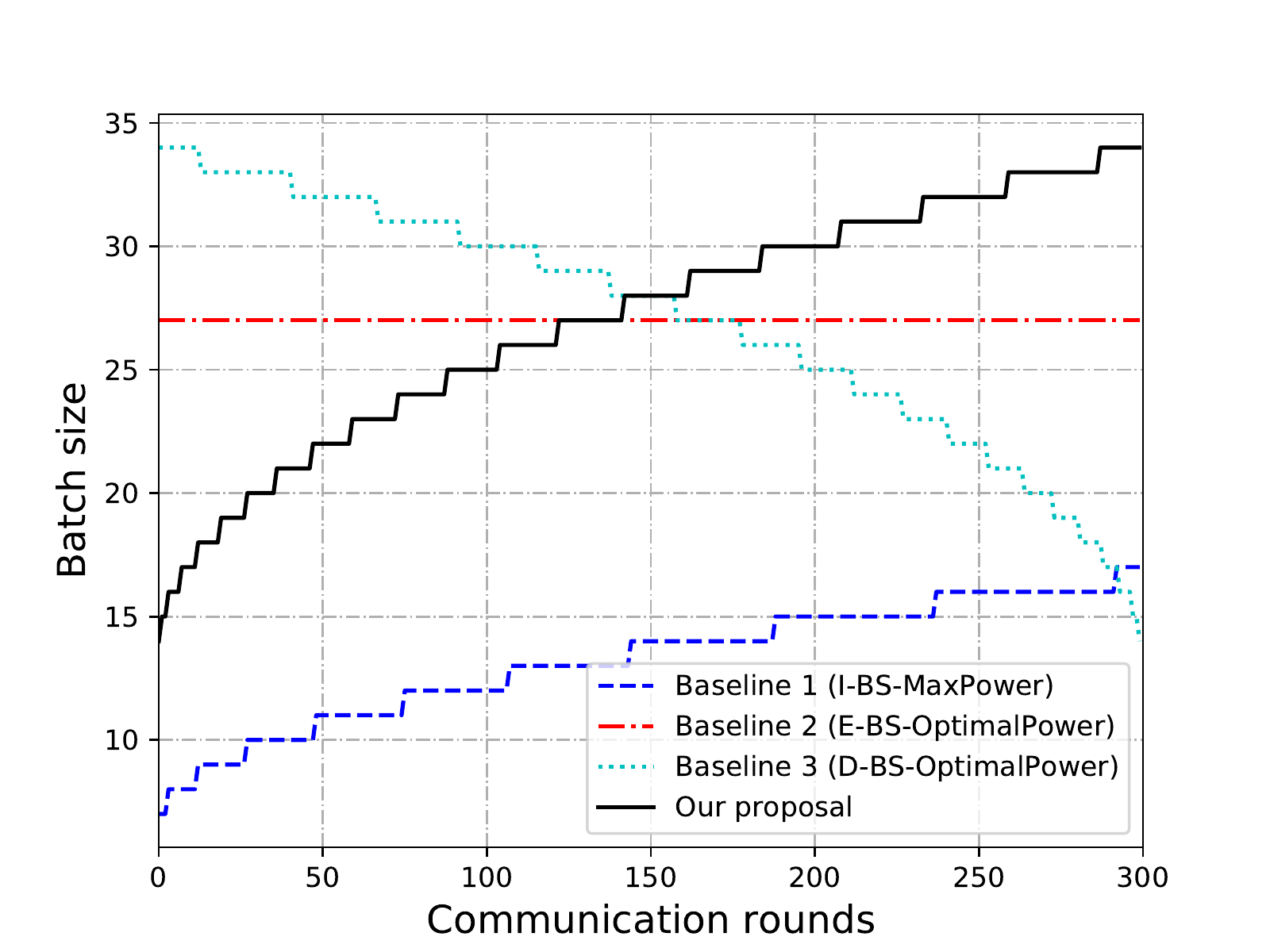}
	\caption{The batch size at each communication round in different schemes in the energy-limited case.}\vspace{-0.5cm}
	\label{fig:fig_batch_round}
\end{figure}

\begin{figure}[!t]
	\centering
	\subfloat[Training loss versus training time]{\includegraphics[width=0.36\textwidth]{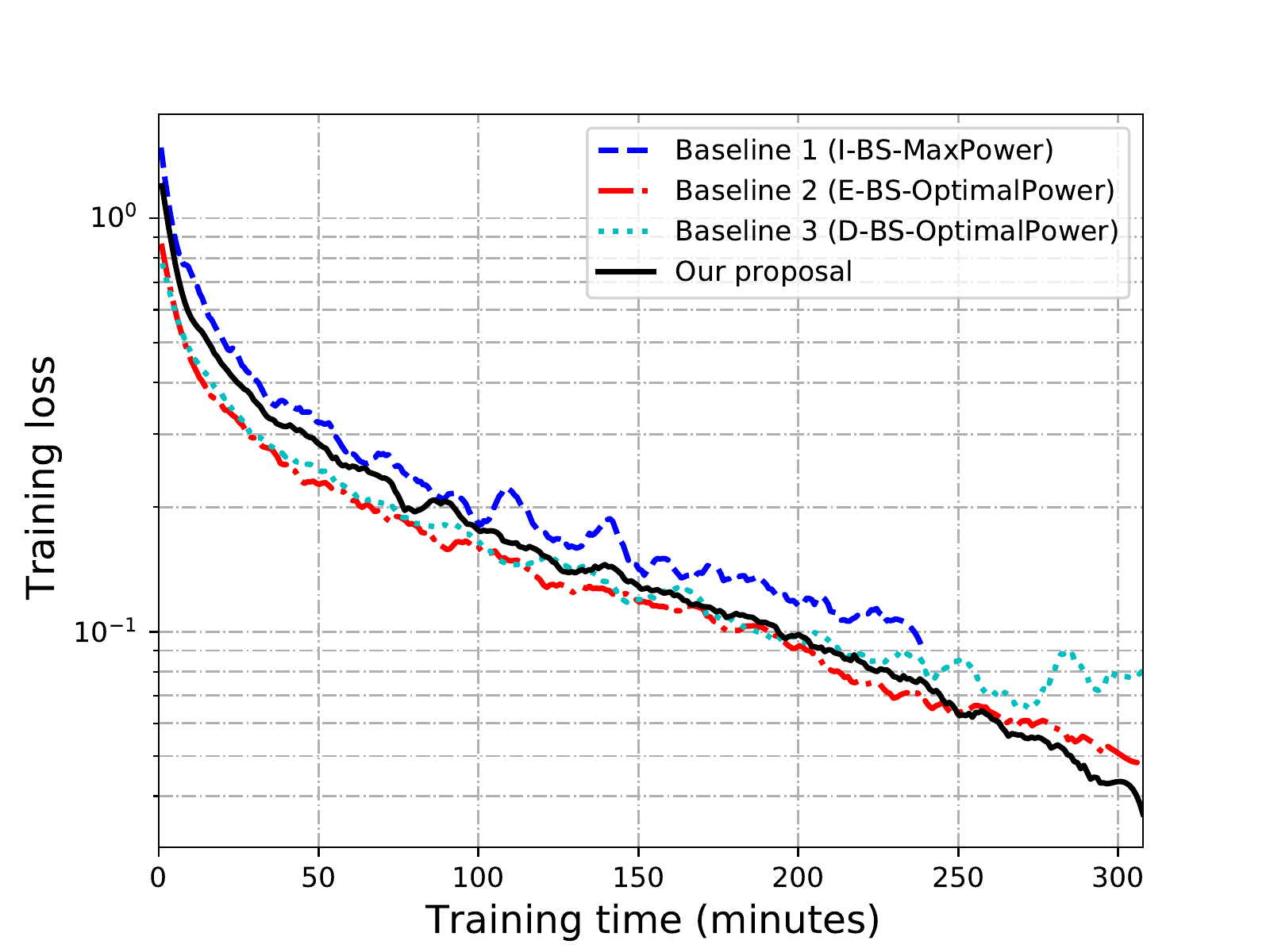}
		\label{fig:fig_training_loss_EL}}
	\\
	\subfloat[Test accuracy versus training time]{\includegraphics[width=0.36\textwidth]{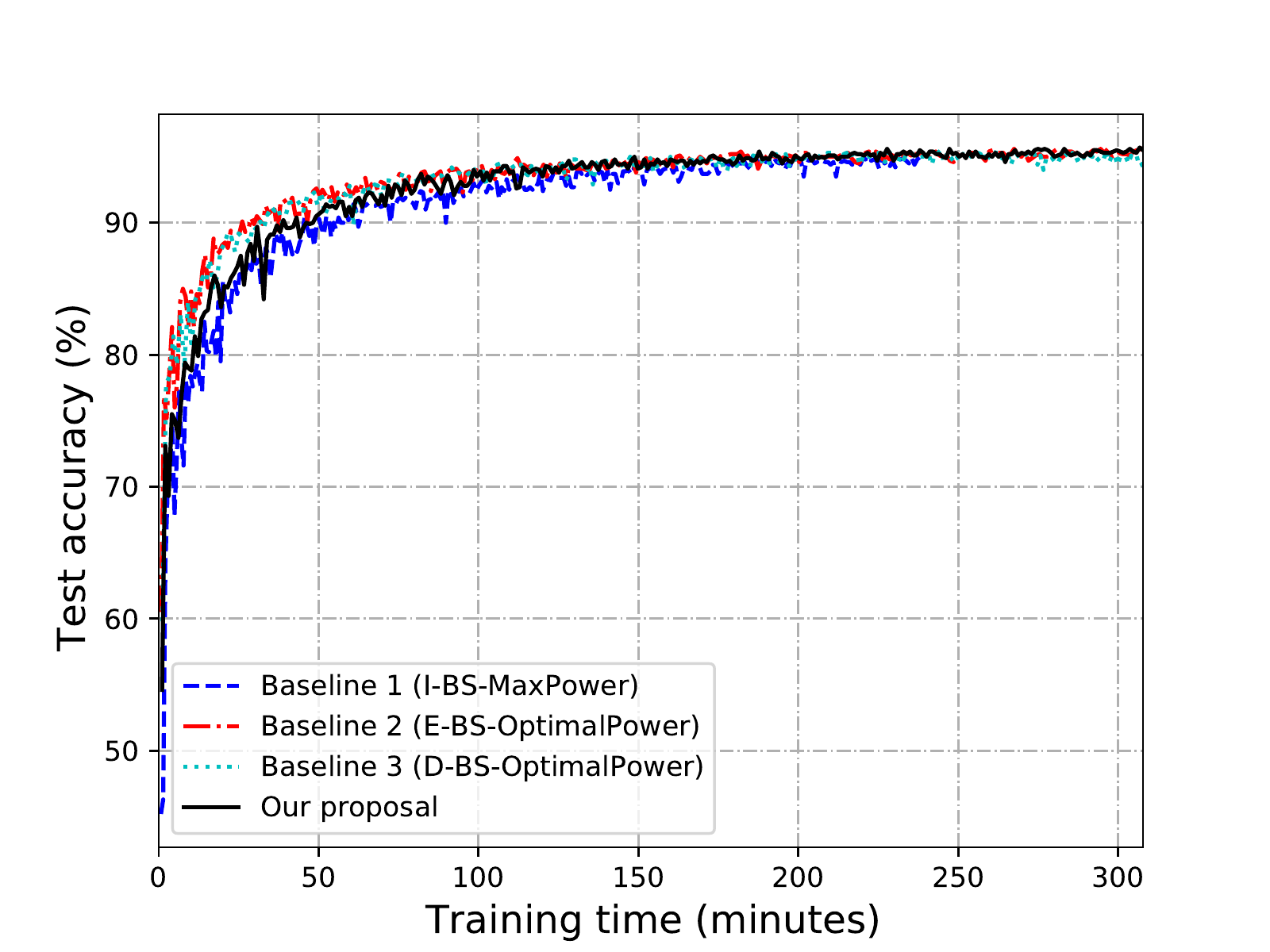}
		\label{fig:fig_accuracy_EL}}
	\caption{Performance comparison among different schemes in the energy-limited case.}\vspace{-0.5cm}
	\label{fig:fig_loss_acc_EL}
\end{figure}

\subsubsection{Selection of hyperparameter $b_{0}$}

In our proposed scheme, we need to choose the hyperparameter $b_{0}$, which determines the batch size at the initial communication round. Since the batch size should be increasing with the communication round, $b_{0}$ should be smaller than the average batch size $b_{\sf avg} = \frac{b_{\sf sum}}{R}$. In this experiment, we evaluate four different values of $b_{0}$, i.e., $b_{0} = 0.25 b_{\sf avg}$, $0.5 b_{\sf avg}$, $0.75 b_{\sf avg}$, and $b_{\sf avg}$. Note that the batch size of each communication round is equal when $b_{0} = b_{\sf avg}$. Fig. \ref{fig:fig_batch_div} depicts the batch size at each communication round under different $b_{0}$. Fig. \ref{fig:fig_training_loss_batch} represents the training speed (loss value versus training time) under different $b_{0}$. It can be observed from Fig. \ref{fig:fig_training_loss_batch} that the training speeds are almost the same when $b_{0} = 0.25 b_{\sf avg}$, $0.5 b_{\sf avg}$ and $0.75 b_{\sf avg}$, but significantly faster than the case when $b_{0} = b_{\sf avg}$ in the later stage of the training process. Therefore, in the following experiments, we will randomly choose $b_{0} = 0.5 b_{\sf avg}$. 

\subsubsection{Optimal communication transmit power}

For exposition, we fix the latency constraint as $T^{\sf max} = 20000$ seconds and evaluate the maximum total sensed samples $b_{\sf sum}^{\ast}$ under different energy constraints $E^{\sf max}$ in Fig. \ref{fig:fig_bsum_energy}.
We compare our proposed scheme with Baseline 1 (I-BS-MaxPower) to validate the necessity of optimizing communication transmit power in Step 1. 
With different energy constraints $E^{\sf max}$ and latency constraints $T^{\sf max}$, the FEEL system can be energy-limited or latency-limited, as explained in Remark \ref{remark:limited}. 
From Fig. \ref{fig:fig_bsum_energy}, we have the following two observations. 
First, in both schemes, when the energy constraint $E^{\sf max}$ is relatively small, the FEEL system will be energy-limited, and the maximum total sensed samples $b_{\sf sum}^{\ast}$ increases with  $E^{\sf max}$; when $E^{\sf max}$ is large enough, the system will be latency-limited, and $b_{\sf sum}^{\ast}$ will no longer increase with $E^{\sf max}$.
Moreover, it can be observed from Fig. \ref{fig:fig_power_EL} that the optimal communication transmit power of each device does not reach the maximum power constraint $P_{{\sf c},k}^{\max}$ in the energy-limited case, which means that optimizing communication transmit power is necessary in an energy-limited FEEL system.
Second, in the energy-limited case, our proposed scheme can lead to much larger $b_{\sf sum}^{\ast}$ than Baseline 1 (I-BS-MaxPower), especially when $E^{\sf max}$ is small. 
These results suggest the superiority of our proposed scheme in allocating communication transmit power.

\subsubsection{Adaptive batch size}

Fig. \ref{fig:fig_loss_acc_EL} shows the comparison of the learning performance among different schemes in the energy-limited case. 
Since all devices transmit with maximum communication power in the latency-limited case, the optimal $b_{\sf sum}^{\ast}$ will be the same for all the schemes in this case. 
Thus, it is not necessary to evaluate the latency-limited case.
Compared with Baseline 1 (I-BS-MaxPower), our proposed scheme can
achieve a faster convergence speed while attaining a higher learning accuracy at the same time.
The reason is that our proposed scheme brings larger $b_{\sf sum}^{\ast}$, and thus a larger batch of samples can be used for SGD at each communication round (see Fig. \ref{fig:fig_batch_round}). 
Moreover, the training process stops before the time constraint in Baseline 1 (I-BS-MaxPower), as it uses maximum communication transmit power at each communication round without allocating the energy properly and runs out of energy early.
Baseline 2 (E-BS-OptimalPower) and Baseline 3 (D-BS-OptimalPower) both obtain the the optimal $b_{\sf sum}^{\ast}$ as same as our proposed scheme, but adopt a different batch size updating rule (see Fig. \ref{fig:fig_batch_round}).
It can be observed that Baseline 2 (E-BS-OptimalPower) achieves a slower convergence speed compared with our proposed scheme, and Baseline 3 (D-BS-OptimalPower) cannot even converge. 
One important observation is that in our proposed scheme the training converges a little slower but speeds up in the later stage of the training process.
This result suggests the superiority of our proposed scheme in adopting adaptive batch size across different communication rounds.

\section{Conclusion}\label{sec:conclusion}

In this paper, we investigated the issue of joint SC$^{2}$ resource allocation for FEEL to enhance the ambient intelligence. 
The significance of the work is two-fold.
First, we have characterized the sensing process in ambient intelligence for a concrete case study, namely human motion recognition, and discovered that sensing with a threshold power is adequate for generating data samples of approximately the same satisfactory quality, which is useful for FEEL.
Second, we have proposed a joint SC$^{2}$ resource allocation scheme that specifies the optimal transmit power and time for communication and sensing, and batch size to be computed at each communication round.
This work opens several future directions for further investigation.
More practical cases could be considered, such as ISAC devices with heterogeneous computation power, and adaptive total communication rounds. 
In addition, there are many different kinds of sensors for ambient intelligence, such as cameras, depth sensors, and radio sensors, which generate data samples of different modalities. 
Thus, how to deal with multi-modality in ambient intelligence together with joint SC$^{2}$ resource allocation for FEEL is also a promising direction.

\bibliographystyle{IEEEtran}
\bibliography{FediSAC}
\end{document}